\newif\iffinal
\finalfalse

\documentclass[sigplan,nonacm,balance=false]{acmart}

\usepackage{amsmath,amsfonts,bm}

\newcommand{\captiona}{{\em (a)}}
\newcommand{\captionb}{{\em (b)}}

\def\Secref#1{Section~\ref{#1}}
\def\eqref#1{equation~\ref{#1}}

\def\1{\bm{1}}

\DeclareMathAlphabet{\mathsfit}{\encodingdefault}{\sfdefault}{m}{sl}
\SetMathAlphabet{\mathsfit}{bold}{\encodingdefault}{\sfdefault}{bx}{n}

\usepackage[noend]{algpseudocode}
\usepackage[normalem]{ulem}
\usepackage{algorithm}
\usepackage{booktabs}
\usepackage{xurl}
\usepackage[capitalise,noabbrev]{cleveref}
\usepackage{makecell}
\usepackage{multirow}
\usepackage{pifont}
\usepackage{subfig}
\newcommand{\papertablestyle}{%
  \footnotesize
  \renewcommand{\arraystretch}{1.05}%
  \setlength{\tabcolsep}{3pt}%
  \setlength{\arrayrulewidth}{0.4pt}%
  \setlength{\heavyrulewidth}{0.8pt}%
  \setlength{\lightrulewidth}{0.4pt}%
}
\newcommand{\tableyes}{\textcolor[rgb]{0,0.60,0.30}{\ding{51}}}
\newcommand{\tableno}{\textcolor[rgb]{0.76,0.16,0.10}{\ding{55}}}
\newcommand{\paperplotfont}{\fontencoding{T1}\fontfamily{DejaVuSans-TLF}\fontseries{m}\fontshape{n}\fontsize{7.2}{8.64}\selectfont}
\DeclareCaptionFont{paperplot}{\paperplotfont}
\usepackage{threeparttable}
\usepackage{tikz}
\usepackage{xspace}
\usepackage{bbm}
\usepackage[most]{tcolorbox}

\theoremstyle{acmplain}

\theoremstyle{acmdefinition}

\newtcolorbox{insightbox}{
  colback=gray!8,
  colframe=gray!40,
  boxrule=0.4pt,
  arc=1.5pt,
  left=6pt,
  right=6pt,
  top=4pt,
  bottom=4pt,
  fontupper=\small
}

\newcommand{\sys}{\textsc{OpWeave}\xspace}

\title{\sys: Flexible Operator Disaggregation for Heterogeneous LLM Serving}

\author{Zikun Li}
\authornote{Zikun Li and Yixuan Mei contributed equally to this work.}
\affiliation{%
  \institution{Carnegie Mellon University}
  \city{}
  \country{}
}
\email{zikunl@andrew.cmu.edu}

\author{Yixuan Mei}
\authornotemark[1]
\affiliation{%
  \institution{Carnegie Mellon University}
  \city{}
  \country{}
}
\email{yixuanm@andrew.cmu.edu}

\author{Shiqi Pan}
\affiliation{%
  \institution{Carnegie Mellon University}
  \city{}
  \country{}
}
\email{shiqip@andrew.cmu.edu}

\author{Zixuan Chen}
\affiliation{%
  \institution{Carnegie Mellon University}
  \city{}
  \country{}
}
\email{zixuanc3@andrew.cmu.edu}

\author{Xiaowen Zhang}
\affiliation{%
  \institution{Carnegie Mellon University}
  \city{}
  \country{}
}
\email{xiaowen5@andrew.cmu.edu}

\author{Mengdi Wu}
\affiliation{%
  \institution{Carnegie Mellon University}
  \city{}
  \country{}
}
\email{mengdiwu@andrew.cmu.edu}

\author{Shuhuai Lin}
\affiliation{%
  \institution{Carnegie Mellon University}
  \city{}
  \country{}
}
\email{shuhuailinux@gmail.com}

\author{Yutong Yang}
\affiliation{%
  \institution{Carnegie Mellon University}
  \city{}
  \country{}
}
\email{aprilytyang@gmail.com}

\author{Zhihao Zhang}
\affiliation{%
  \institution{Carnegie Mellon University}
  \city{}
  \country{}
}
\email{zhihaoz3@andrew.cmu.edu}

\author{Xupeng Miao}
\authornote{Work done at Purdue.}
\affiliation{%
  \institution{Peking University}
  \city{}
  \country{}
}
\email{xupeng.miao@pku.edu.cn}

\author{Rashmi Vinayak}
\affiliation{%
  \institution{Carnegie Mellon University and Google}
  \city{}
  \country{}
}
\email{rvinayak@andrew.cmu.edu}

\author{Zhihao Jia}
\affiliation{%
  \institution{Carnegie Mellon University}
  \city{}
  \country{}
}
\email{zhihao@cmu.edu}

\renewcommand{\shortauthors}{Li et al.}
\begin{document}

\begin{abstract}
LLM serving systems increasingly disaggregate inference into finer-grained stages, with recent approaches separating attention from FFN or MoE execution during decode. This \emph{operator-level disaggregated serving} (ODS) can improve hardware matching and enable independent scaling, particularly across heterogeneous devices. However, existing systems fix operator boundaries and lack a unified characterization of when disaggregation reduces serving cost. We present \sys, an end-to-end framework for heterogeneous ODS. \sys provides an analytical cost model that bounds the gains of homogeneous and heterogeneous ODS over colocated serving. It jointly optimizes operator partitioning and deployment configuration through a regularity-aware planner that keeps the search tractable even for hybrid-attention models. A vLLM-based runtime executes the synthesized plans with flexible operator stages across heterogeneous device groups. In our evaluation, \sys reduces serving cost by up to $1.78\times$ on homogeneous and $1.89\times$ on heterogeneous GPU clusters relative to the best feasible baseline, while meeting latency SLOs.
\end{abstract}

\maketitle

\section{Introduction}
\label{sec:introduction}

\begin{figure}[t]
    \centering
    \includegraphics[width=\columnwidth]{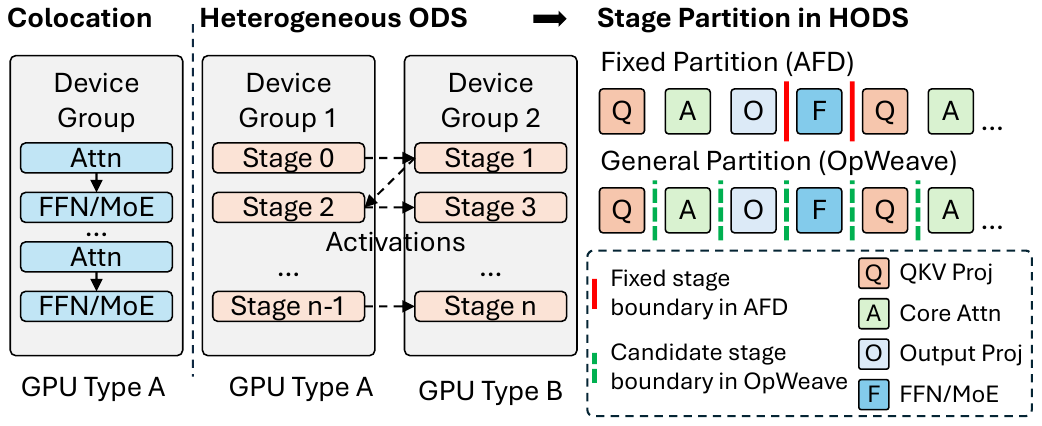}
    \caption{Colocated serving versus heterogeneous operator-level disaggregated serving (ODS). AFD fixes stage boundaries at attention--FFN interfaces, whereas \sys supports general operator partitions across heterogeneous device groups.}
    \label{fig:ods_vs_colocation}
\end{figure}

Large language model (LLM) inference accounts for a substantial share of AI infrastructure spending and requires significant investment in compute~\cite{menlo2025midyear,deloitte2025infrastructure,noffsinger2025cost_of_compute}. At this scale, even modest improvements in throughput and hardware utilization, or reductions in per-token cost, can yield substantial savings. Improving LLM serving efficiency has therefore become a central systems challenge~\cite{yu2022orca,kwon2023pagedattention,agrawal2024sarathi,patel2024splitwise,zhong2024distserve}.

Modern serving systems increasingly separate inference phases to optimize and scale them independently. 
Most existing approaches disaggregate coarse-grained phases, particularly prefill and decode~\citep{patel2024splitwise,zhong2024distserve,singh2025epd,dong2025hydrainfer,lohmeyer2025gke_inference_gateway_ga,gangasani2026aws_llmd_disaggregated_inference}. However, these approaches generally retain decode as a single execution unit, despite substantial differences in the resource demands of its constituent operators. Recent systems extend disaggregation into the decode phase by separating attention from feed-forward network (FFN) or mixture-of-experts (MoE) execution~\citep{aubrey2026groq3lpx,zhu2025megascaleinfer,stepfun2025step3,song2026attention_ffn_ratios,liu2026afd_challenges}. This development motivates our focus on operator-level disaggregated serving.

\emph{Operator-level disaggregated serving} (ODS) partitions decode computation into separately executed operator stages, such as attention and FFN or MoE stages. \emph{Heterogeneous ODS} further assigns these stages to different device types, as illustrated in \cref{fig:ods_vs_colocation}. This design addresses two sources of inefficiency in conventional serving engines~\citep{yu2022orca,kwon2023pagedattention,zheng2023sglang,agrawal2024sarathi}. First, colocating operators can lead to \emph{poor hardware matching}: decode attention is often limited by memory bandwidth and underutilizes tensor cores, while FFN and MoE operators rely heavily on matrix multiplications and benefit from high arithmetic throughput~\citep{zhu2024nanoflow,zhu2025megascaleinfer,stepfun2025step3}. Second, colocation introduces \emph{memory-capacity coupling}: as sequence length increases, the growing key–value (KV) cache limits the feasible batch size for the entire workload, potentially constraining FFN and MoE throughput.

Heterogeneous ODS mitigates these inefficiencies by pipelining operator stages across groups of heterogeneous GPUs~\citep{he2024fastdecode,zhu2025megascaleinfer,stepfun2025step3,song2026attention_ffn_ratios,liu2026afd_challenges}. First, it improves \emph{hardware matching} by assigning each stage to devices suited to its resource requirements. Second, it enables \emph{independent scaling}, allowing FFN and MoE stages to use batch sizes that are not directly constrained by the KV-cache capacity of an individual attention stage. Together, these capabilities can improve decode efficiency and reduce serving cost.

Despite this potential, realizing these benefits requires addressing three challenges.

\begin{figure}[t]
    \centering
    \subfloat[\label{fig:gemma_tensor_core_utilization_b200}]{%
        \includegraphics[width=0.48\columnwidth]{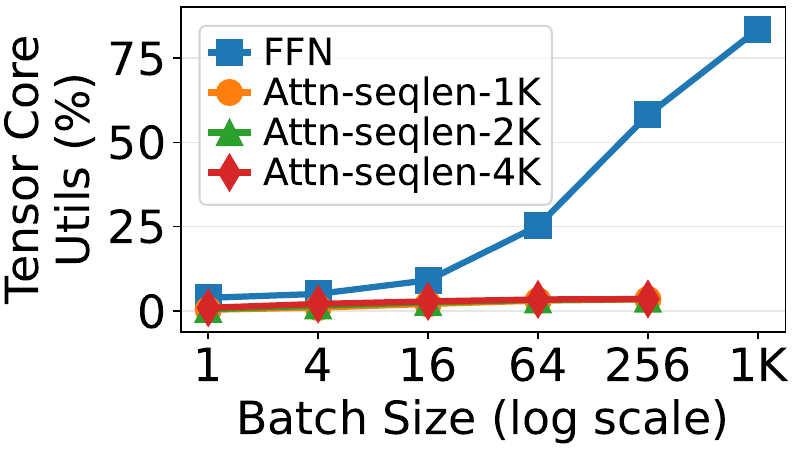}%
    }
    \hfill
    \subfloat[\label{fig:h100_gemma3_max_batch_size_vs_seqlen}]{%
        \includegraphics[width=0.48\columnwidth]{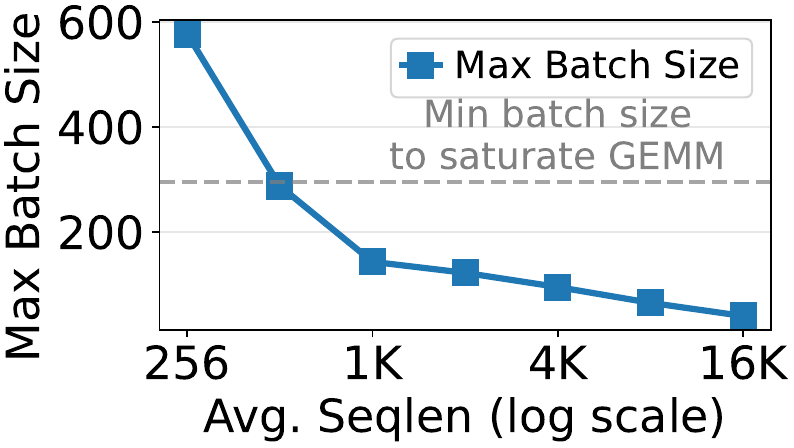}%
    }
    \caption{Decode-phase bottlenecks of Gemma-3-27B motivating ODS. \captiona~Tensor-core utilization of attention and FFN operators across batch sizes (B200): attention remains underutilized while batching improves FFN utilization. \captionb~Maximum feasible batch size across sequence lengths (H100): KV-cache growth constrains batching.}
    \label{fig:decode_imbalance}
\end{figure}

\paragraph{Theory gap.}
Existing work provides empirical evidence and analytical models for specific ODS designs~\citep{zhu2025megascaleinfer,stepfun2025step3,song2026attention_ffn_ratios,liu2026afd_challenges}. A unified analysis is needed to determine when ODS reduces cost relative to colocated serving, establish bounds on the benefits of homogeneous and heterogeneous ODS, and characterize how these benefits depend on sequence length, model architecture, and hardware characteristics.

\paragraph{Planning gap.}
Existing systems optimize deployments around fixed operator boundaries, such as attention--FFN or attention--MoE splits~\citep{zhu2025megascaleinfer,stepfun2025step3}. These fixed boundaries constrain joint optimization of operator partitioning, hardware assignment, parallelism, and batching. A general formulation is needed to synthesize efficient deployment plans across this broader design space, particularly for hybrid-attention models whose operator structures vary across layers.

\paragraph{Runtime gap.}
Existing ODS runtimes organize execution and communication around specific attention--FFN or attention--MoE splits~\citep{zhu2025megascaleinfer,stepfun2025step3}. Executing synthesized plans with different operator boundaries and pipeline schedules requires flexible operator staging, efficient pipeline orchestration, and low-overhead communication across heterogeneous device groups.

To address these challenges, we present \sys, an end-to-end framework for heterogeneous ODS that integrates theoretical analysis, deployment planning, and runtime execution.

First, \sys uses an analytical cost model for colocated serving, homogeneous ODS, and heterogeneous ODS. The analysis characterizes how cost reductions depend on sequence length, model architecture, and hardware, and upper-bounds the gains of disaggregation and heterogeneous hardware assignment. Under this model, \emph{core-attention disaggregation} (CAD) realizes the idealized ODS cost when its stages are balanced and communication is fully overlapped. The analysis also identifies two limitations of CAD: hardware and workload constraints can prevent stage balance, and layerwise variation in hybrid-attention models can introduce pipeline imbalance under a fixed two-stage partition.

Second, \sys formulates heterogeneous ODS planning as a joint optimization problem over operator partitioning, hardware assignment, parallelism, batching, and pipeline orchestration. To make this combinatorial search tractable, we introduce the \emph{partition block}: the smallest repeating, layer-aligned operator pattern in a model, used as the unit of plan construction and reuse. Building on this abstraction, \sys develops a \emph{regularity-aware} planner that reduces the search space by reusing partitioning decisions across layers with the same high-level operator structure, even when attention mechanisms differ. Together they support operator partitions beyond fixed attention--FFN or attention--MoE splits.

\begin{figure}[t]
    \centering
    \includegraphics[width=0.9\columnwidth]{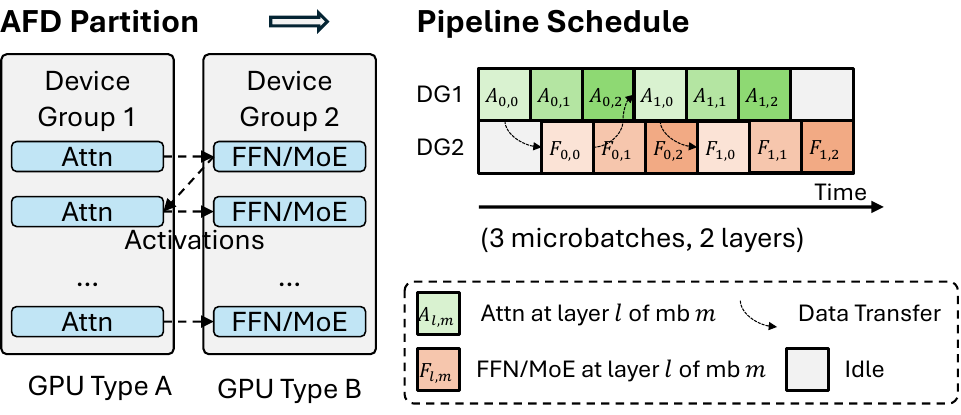}
    \caption{Pipelined attention--FFN disaggregation (AFD).}
    \label{fig:afd_pipeline}
\end{figure}

Third, \sys executes synthesized serving plans through a distributed runtime built on vLLM. The runtime's staged engines support flexible operator partitions and pipeline schedules across heterogeneous device groups. To reduce host orchestration overhead, each worker submits its computation and communication sequence in one native call per decoding iteration. Low-overhead communication further reduces transfer costs.

Finally, we evaluate \sys across representative models, hardware configurations, and long-context workloads. On homogeneous H100 clusters, \sys reduces serving cost by up to $1.78\times$ compared to the best existing approaches while attaining time-per-output-token (TPOT) service-level objectives (SLOs). Compared with attention--FFN disaggregation (AFD), \sys plans require fewer pipeline stages (20.7 versus 124 on average for Gemma-3-27B) and reduce inter-node data transfer per output token by up to $22.8\times$; the resulting latency reduction lets \sys meet stringent TPOT SLOs that AFD cannot satisfy. \sys also keeps hardware executing the model over 80\% of the time, whereas AFD leaves it idle roughly half the time. On heterogeneous clusters combining H100 with L40S or A100 GPUs, heterogeneous \sys reduces serving cost by up to $1.89\times$ compared to the best non-\sys baseline and by up to $21.8\%$ compared to homogeneous \sys.
\section{Background and Motivation}
\label{sec:background}

\paragraph{Transformer Inference}
Transformer-based LLMs stack repeated layers of attention and feed-forward computation, implemented as FFN or MoE blocks. Inference consists of two phases with distinct computational profiles: prefill processes the input context in parallel and is typically compute-bound, whereas decode generates one token per autoregressive iteration. This contrast motivates prefill-decode disaggregation, which places the two phases on separate resources for independent optimization~\cite{patel2024splitwise,zhong2024distserve}. For generation-heavy workloads, such as reasoning tasks that produce long output sequences, decode often dominates latency and serving cost.

\begin{figure}[t]
    \centering
    \includegraphics[width=\columnwidth]{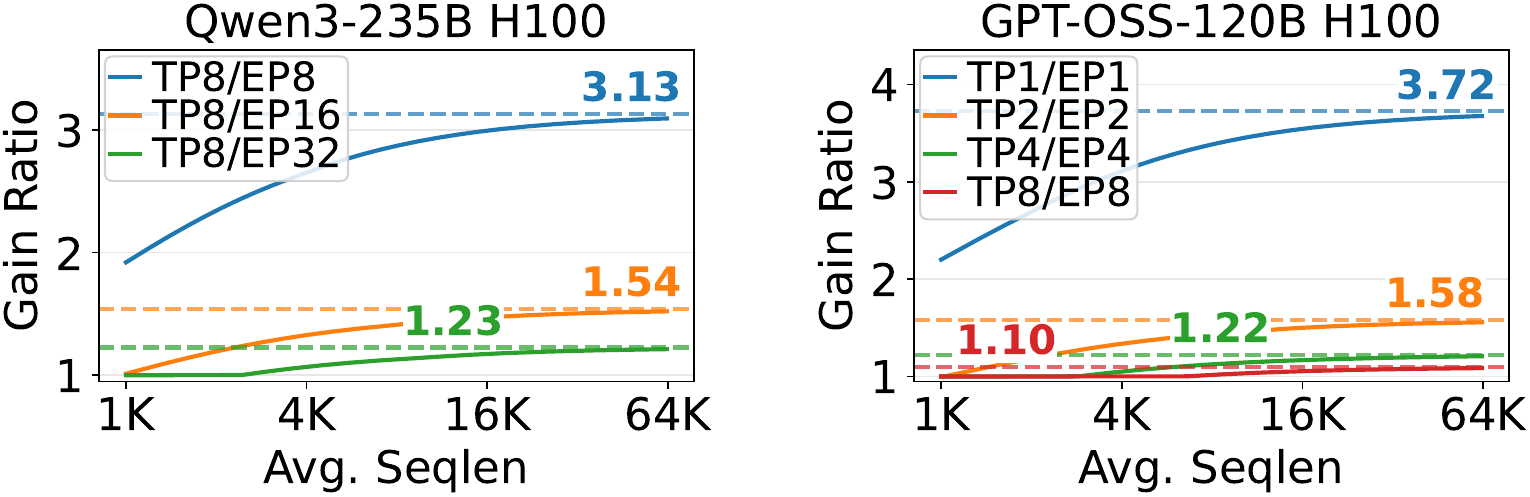}
    \caption{Theoretical cost gain of homogeneous ODS over colocated serving. Solid curves show gains for Qwen3-235B and GPT-OSS-120B on H100 under different attention-TP/MoE-EP configurations; dashed lines show upper bounds.}
    \label{fig:homo_vs_coloc}
\end{figure}

\paragraph{Decode-Phase Tensions}
Serving systems batch concurrent requests to amortize model-weight accesses and improve hardware utilization, particularly for GEMM-heavy FFN and MoE operators. However, attention and FFN/MoE benefit differently from batching. Attention remains memory-band\-width-bound and exhibits low tensor-core utilization even as batch size increases, as shown in Figure~\ref{fig:gemma_tensor_core_utilization_b200}. Moreover, as context length grows, the attention key-value (KV) cache reduces the maximum feasible batch size, as shown in Figure~\ref{fig:h100_gemma3_max_batch_size_vs_seqlen}, preventing FFN and MoE operators from reaching the batch sizes needed for efficient execution.

\paragraph{Attention--FFN Disaggregation}
Prior heterogeneous ODS systems, including Step-3 and MegaScale-Infer, primarily adopt a fixed attention--FFN disaggregation (AFD) design rather than searching over general operator partitions~\cite{stepfun2025step3,zhu2025megascaleinfer}. The attention stage executes the full attention module, including the QKV and output projections, while the FFN/MoE stage executes feed-forward or expert computation. Because attention and FFN/MoE alternate within each Transformer layer, hidden states must be transferred between the device groups at each stage boundary, as shown in Figure~\ref{fig:afd_pipeline}. These systems divide batches into microbatches and pipeline their execution to keep both stages busy while overlapping inter-stage communication with computation. Consequently, performance depends strongly on stage balance: latency mismatches create pipeline bubbles and reduce utilization. Although prior systems demonstrate the feasibility of AFD, their fixed partitions leave unresolved when heterogeneous ODS reduces serving cost and what determines its benefit, the question we address next through theoretical analysis.

\section{Theoretical Analysis}
\label{sec:theory_analysis}
\label{sec:theory}

\subsection{Analytical Setup}
\label{sec:analytical_setup}

\paragraph{Scope and assumptions.}
We analyze steady-state decode serving with memory-bound attention and GEMM-dominated projections, FFN, and MoE computation. Each GPU type has unit-time monetary cost $c$, memory bandwidth $\beta$, memory capacity $M$, and peak compute throughput $F$; we minimize GPU cost per decoded token. We model attention execution time as memory traffic divided by memory bandwidth, and GEMM execution time as FLOP count divided by peak GPU compute throughput only in the compute-bound regime. We assume disaggregation enables GEMM batches large enough for compute-bound execution, and that microbatching and pipelining hide inter-device communication. Appendices~\ref{app:analytical-model-details} and~\ref{app:execution-regimes} give the operator-level roofline model and formal regime assumptions.

\begin{figure}[t]
    \centering
    \includegraphics[width=\columnwidth]{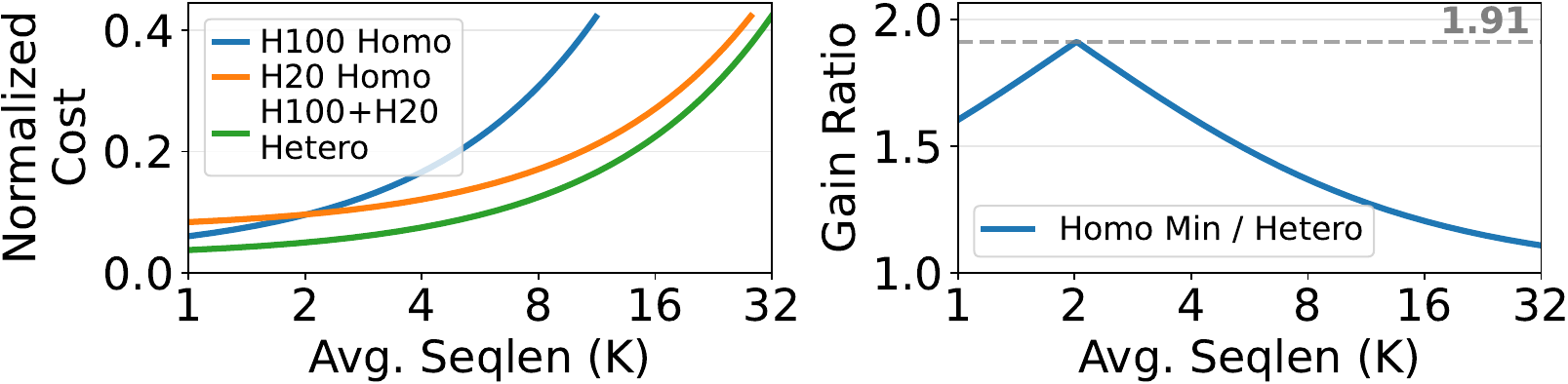}
    \caption{Heterogeneous vs.\ homogeneous ODS for Qwen3-235B on H100 and H20 GPUs. The heterogeneous setup runs core attention on H20 and GEMM-dominated operators on H100. \emph{Left:} normalized cost per token. \emph{Right:} cost gain over the best homogeneous deployment.}
    \label{fig:hetero_vs_homo}
\end{figure}

\paragraph{Cost model.}
Let $m_{\mathrm{kv}}(s)$ be the KV-cache memory per request at average sequence length $s$. Let $P_{\mathrm{act}}$ be the effective number of GEMM-side parameters activated per request after model-parallel sharding, and let $M_{\mathrm{weights}}$ be the resident model-weight footprint on the modeled device group, including inactive MoE experts. For GPU type $j$, define $\alpha_j \triangleq c_j/\beta_j$ and $\gamma_j \triangleq c_j/F_j$ as monetary cost per byte of memory traffic and per FLOP, respectively. Under the regime assumptions above, homogeneous ODS on type $j$ has cost
\begin{equation}
\label{eq:cpt_homo_disagg}
\mathrm{CPT}^{\mathrm{hom}}_j(s)
=
\alpha_j m_{\mathrm{kv}}(s)
+
2\gamma_j P_{\mathrm{act}},
\end{equation}
while heterogeneous ODS with attention on GPU type $j$ and GEMM-dominated operators on type $k$ has cost
\begin{equation}
\label{eq:cpt_hetero_disagg}
\mathrm{CPT}^{\mathrm{het}}_{j,k}(s)
=
\alpha_j m_{\mathrm{kv}}(s)
+
2\gamma_k P_{\mathrm{act}}.
\end{equation}
For colocated serving on type $j$ with memory capacity $M$, define the continuous relaxation $\bar b(s)$ and the exact integer batch limit $b_{\max}(s)$ as
\begin{equation}
\label{eq:batch_limit}
\bar b(s)
\triangleq
\frac{M-M_{\mathrm{weights}}}{m_{\mathrm{kv}}(s)}
,
\qquad
b_{\max}(s)=\lfloor\bar b(s)\rfloor.
\end{equation}
For feasible sequence lengths, the minimum colocated cost under the continuous batch-size relaxation is
\begin{equation}
\label{eq:cpt_coloc}
\mathrm{CPT}^{\mathrm{coloc}}_j(s)
=
\alpha_j m_{\mathrm{kv}}(s)
+
\max\!\left(
\frac{\alpha_j M_{\mathrm{weights}}}{\bar b(s)},\,
2\gamma_j P_{\mathrm{act}}
\right).
\end{equation}
The largest relaxed batch $\bar b(s)$ attains this minimum because cost is non-increasing in batch size. The maximum captures whether GEMM is limited by weight loading or compute.

\paragraph{GEMM threshold.}
For local batch size $b$, Eqs.~\ref{eq:cpt_homo_disagg}--\ref{eq:cpt_coloc} use the approximations that attention traffic scales as $b\,m_{\mathrm{kv}}(s)$, active GEMM work as $2bP_{\mathrm{act}}$, and GEMM weight traffic as $M_{\mathrm{weights}}$. Equating GEMM weight-loading time and compute time gives the threshold
$b^* = \frac{M_{\mathrm{weights}}F}{2P_{\mathrm{act}}\beta}$,
which is the batch size at which GEMM becomes compute-bound. Appendix~\ref{app:cost-expression-details} derives the cost expressions, and Appendix~\ref{app:scaling-threshold-details} states the scaling approximations and threshold derivation in full.

\subsection{Homogeneous Disaggregation vs.\ Colocation}
\label{sec:homo_vs_coloc}

We first analyze when homogeneous operator-level disaggregation reduces cost relative to colocation, and how that gain depends on sequence length.
We fix the GPU type and omit its index.
We restrict the comparison to sequence lengths for which colocation is feasible, i.e., $b_{\max}(s) \ge 1$, equivalently $\bar b(s) \ge 1$, and assume that $m_{\mathrm{kv}}(s)$ is strictly increasing on this domain.
Define the \emph{cost gain} of homogeneous disaggregation over colocation as $G_{\mathrm{hom}}(s) \triangleq \frac{\mathrm{CPT}^{\mathrm{coloc}}(s)}{\mathrm{CPT}^{\mathrm{hom}}(s)}$.

\begin{insightbox}
\begin{theorem}[Cost gain of homogeneous disaggregation over colocation]
\label{thm:homo_vs_coloc}
Consider a homogeneous GPU setting with parameters $(c, \beta, M, F)$. Suppose attention is memory-bound and adopt the scaling approximations from Section~\ref{sec:analytical_setup}. Let $b^*$ be the GEMM compute-bound threshold (Section~\ref{sec:analytical_setup}) and let $\bar b(s)$ be the relaxed feasible batch size defined in Eq.~\ref{eq:batch_limit}. Then, for every feasible sequence length $s$ with $b_{\max}(s) \ge 1$:
\begin{enumerate}
    \item \emph{(No gain at short sequences.)} If $\bar b(s) \geq b^*$, then $G_{\mathrm{hom}}(s) = 1$.
    \item \emph{(Strictly increasing gain.)} On the feasible region where $\bar b(s) < b^*$, $G_{\mathrm{hom}}(s) > 1$ and $G_{\mathrm{hom}}$ is strictly increasing in $s$.
    \item \emph{(Bounded gain.)} The gain satisfies
    \begin{equation}
        \label{eq:homo_bound}
        1 \;\leq\; G_{\mathrm{hom}}(s) \;<\; \frac{M}{M - M_{\mathrm{weights}}}.
    \end{equation}
\end{enumerate}
\end{theorem}
\end{insightbox}

Homogeneous disaggregation helps only after KV-cache growth pushes the colocated feasible batch size below the GEMM compute-bound threshold. Beyond that point, the gain increases monotonically with sequence length but remains bounded by $M/(M-M_{\mathrm{weights}})$. Disaggregation improves efficiency by decoupling GEMM batching from KV-cache memory pressure, so its benefit is fundamentally limited by how much device memory is already consumed by model weights. Figure~\ref{fig:homo_vs_coloc} shows cost gains and upper bounds calculated from our theoretical model for Qwen3-235B and GPT-OSS-120B on H100 under different parallelism configurations. See Appendix~\ref{app:proof_homo} for the proof.

\subsection{Heterogeneous vs.\ Homogeneous Disaggregation}
\label{sec:hetero_vs_homo}

Optimizing Eqs.~\ref{eq:cpt_homo_disagg}--\ref{eq:cpt_hetero_disagg} over all available GPU types $j$ and $k$ gives
\begin{align}
    \mathrm{CPT}^{\mathrm{hom}}_{\star}(s)
    &= \min_j \mathrm{CPT}^{\mathrm{hom}}_{j}(s),
    \label{eq:cpt_hom_star} \\
    \mathrm{CPT}^{\mathrm{het}}_{\star}(s)
    &= \min_j(\alpha_j)\, m_{\mathrm{kv}}(s) + \min_k(\gamma_k)\, 2P_{\mathrm{act}}.
    \label{eq:cpt_het_star}
\end{align}
The optimized heterogeneous deployment assigns attention to the GPU type with the lowest cost per byte and GEMM-dominated operators to the type with the lowest cost per FLOP; the homogeneous deployment uses one type for both.

Define $G_{\mathrm{het}}(s) \triangleq \frac{\mathrm{CPT}^{\mathrm{hom}}_{\star}(s)}{\mathrm{CPT}^{\mathrm{het}}_{\star}(s)}$ as the gain over homogeneous disaggregation. For the theorem below, restrict these minima to $j,k \in \{1,2\}$ and define $\eta \triangleq (\beta_1 F_2)/(\beta_2 F_1)$ as the hardware ratio.
Assume $\eta \ge 1$ without loss of generality. Let $\mathcal{I}$ be a sequence-length interval where all compared ODS deployments are feasible and $m_{\mathrm{kv}}(s)$ is continuous and strictly increasing.

\begin{insightbox}
\begin{theorem}[Cost gain of heterogeneous over homogeneous disaggregation]
\label{thm:hetero_vs_homo}
Consider two GPU types with parameters $(\beta_1,F_1)$ and $(\beta_2,F_2)$, and respective unit-time costs $c_1$ and $c_2$. Adopt the scaling approximations of Section~\ref{sec:analytical_setup}. Then, over the feasible interval $\mathcal{I}$:
\begin{enumerate}
    \item \emph{(Non-negative gain.)} For every $s \in \mathcal{I}$, $G_{\mathrm{het}}(s) \ge 1$.
    \item \emph{(No gain under dominance.)} If one GPU type dominates the other on both cost-efficiency metrics, i.e., $\alpha_j \le \alpha_k$ and $\gamma_j \le \gamma_k$ for some $j \neq k$, then $G_{\mathrm{het}}(s)=1$ for every $s \in \mathcal{I}$.
    \item \emph{(Unimodal and bounded gain.)} Otherwise, under the labeling $\eta \ge 1$, we have $\alpha_1 < \alpha_2$ and $\gamma_2 < \gamma_1$. Define the homogeneous-cost crossing
    \[
        x^* \triangleq \frac{2P_{\mathrm{act}}(\gamma_1-\gamma_2)}{\alpha_2-\alpha_1}.
    \]
    As a function of $x=m_{\mathrm{kv}}(s)$, the gain strictly increases for $x<x^*$ and strictly decreases for $x>x^*$. If $x^* \in m_{\mathrm{kv}}(\mathcal{I})$, the gain therefore has a unique peak at the corresponding $s^* \in \mathcal{I}$; otherwise, it is monotone over $\mathcal{I}$. Moreover, for every $s \in \mathcal{I}$,
    \begin{equation}
        \label{eq:hetero_bound}
        G_{\mathrm{het}}(s) \;\le\; \frac{1+\sqrt{\eta}}{2}.
    \end{equation}
\end{enumerate}
\end{theorem}
\end{insightbox}

The theorem shows that heterogeneous assignment helps only when the two GPU types trade off bandwidth efficiency against compute efficiency, rather than one dominating on both dimensions. When the homogeneous-cost crossing lies in the feasible interval, heterogeneity helps most there: at shorter sequences the best homogeneous GPU already matches the compute side well, while at longer sequences the best homogeneous GPU already matches the bandwidth side well. If the crossing lies outside the feasible interval, only the increasing or decreasing branch appears in that interval. Figure~\ref{fig:hetero_vs_homo} shows normalized costs per token and the gain of heterogeneous disaggregation over the best homogeneous deployment, calculated from our theoretical model for Qwen3-235B using H100 and H20 GPUs. The full proof is given in Appendix~\ref{app:proof_hetero}.

\subsection{From Bounds to Plans: CAD and Its Limits}
\label{sec:cad_limits}

The bounds above assume that attention and the remaining operators run on hardware suited to their  bottlenecks and that communication is fully hidden by computation. Core-Attention Disaggregation (CAD) implements this separation by placing core attention on one device group and the remaining projection, FFN, and MoE GEMMs on another. Under the assumptions in Section~\ref{sec:analytical_setup}, attaining the corresponding idealized ODS cost requires balanced stage latencies and sufficient microbatching to hide communication (Appendix~\ref{app:cad_achievability}).

\begin{figure}[t]
    \centering
    \includegraphics[width=\columnwidth]{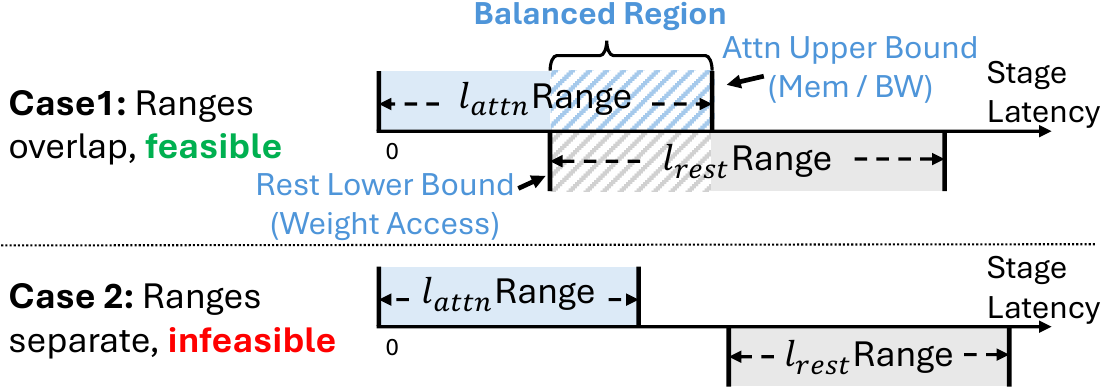}
    \caption{Feasibility of balanced CAD. \emph{Case~1:} the attention and rest-of-model latency ranges overlap, admitting a balanced operating point. \emph{Case~2:} the ranges are disjoint, so pipeline bubbles are unavoidable.}
    \label{fig:cad_feasibility}
\end{figure}

The first limitation, however, is whether stage balance is feasible. GPU memory must accommodate the KV caches of all resident microbatches and attention layers, constraining each attention operation's KV-cache size and hence its maximum latency under our memory-bound model. GEMM-based operators, meanwhile, must load their weights from HBM, imposing a latency floor that our compute-bound cost model does not capture. If this floor exceeds the maximum feasible attention latency, the two stages' feasible latency ranges do not overlap, and the attention stage remains idle for part of every pipeline cycle. Higher memory bandwidth on the GEMM devices lowers this floor and can therefore make stage balance feasible. Figure~\ref{fig:cad_feasibility} illustrates both cases, and Appendix~\ref{app:cad_feasibility_condition} derives the overlap condition for uniform-layer models.

The NVIDIA Vera Rubin and Groq 3 LPX design illustrates this principle: it retains the KV cache and decode attention on GPUs while placing FFN or MoE weights in LPX's SRAM~\citep{aubrey2026groq3lpx}. SRAM residency reduces weight-loading latency, lowering the FFN/MoE stage's latency floor and potentially enabling its feasible latency range to overlap with the attention stage's. Although this design uses AFD rather than CAD, the same two-stage balance principle applies.

\begin{figure}[t]
    \centering
    \includegraphics[width=\columnwidth]{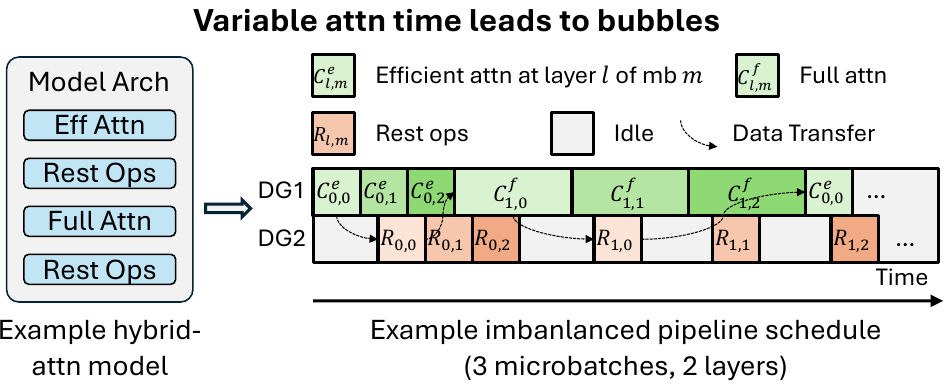}
    \caption{CAD pipeline schedule for a hybrid-attention model. A fixed rest-of-model stage cannot balance both efficient- and full-attention layers, leaving idle pipeline slots.}
    \label{fig:imbalanced_cad}
\end{figure}

The second limitation is layerwise variation. Hybrid architectures commonly
mix full attention with sliding-window, linear, or recurrent sequence
operators~\citep{gemma3,gpt_oss,qwen3_next,team2025kimi,ai2_olmo_hybrid_2026}.
Full attention loads KV cache for the entire context, whereas efficient
operators access a bounded window or compact recurrent state. CAD places both on the same attention
device group, but the rest-of-model stage has one fixed latency for a given
microbatch size and cannot match both layer types. This mismatch leaves one
stage waiting for the other at some layers, creating the pipeline bubbles
shown in Figure~\ref{fig:imbalanced_cad}.

\section{Serving-Plan Formulation}
\label{sec:formulation}

CAD's limitations motivate searching over general operator partitions rather
than fixed module splits.  Our formulation jointly determines how operators are
partitioned into stages, which device-group types execute them with what
hardware and parallelism, and how resident batches and microbatches are sized.
It evaluates each plan using a steady-state latency model that accounts for
both stage execution and inter-stage activation transfers, and minimizes
serving cost subject to memory and TPOT constraints.

\paragraph{Partition-block abstraction.}
The key challenge is to capture cross-layer heterogeneity without searching an
unconstrained partition of the entire model.  We address it with a
partition block, the smallest layer-aligned operator pattern that
repeats across the model.  A partition block may contain one layer in a uniform
transformer or span a full period of full-attention, sliding-window, and other
layer types in a hybrid-attention model.  Within a block of
$N_{\mathrm{block}}$ ordered operators, the planner chooses a contiguous stage
template $\mathbf{q}=(q_0,\ldots,q_S)$, where
$0=q_0<\cdots<q_S=N_{\mathrm{block}}$ and stage~$j$ contains operators
$q_{j-1}+1,\ldots,q_j$.  Stages may span layer boundaries, allowing the
planner to balance heterogeneous operators jointly across layers.  Reusing the
resulting template across all partition blocks preserves the model's regularity
and reduces the partition search to one repeating unit.  Each stage then serves
as the basic unit of placement and pipeline execution;
Appendix~\ref{sec:model-partition} provides the formal DAG, ordering, and block
definitions.

\paragraph{Structured placement and batching.}
Given the stage template, the planner determines where each stage executes
and with what resources.  We represent the deployment using $K$ ordered
device-group types.  Type~$m$ has configuration
$(h_m,\pi_m)$, consisting of a GPU type and a supported parallelism strategy
such as tensor or expert parallelism, and may be instantiated by multiple
\textbf{replicas}.  All replicas of a type execute the same assigned stage
positions while processing disjoint subsets of the global batch.  To obtain a
reusable pipeline structure, we require $S=kK$ for some positive integer~$k$
and assign stage~$j$ using the structured round-robin map
$\sigma(j)=((j-1)\bmod K)+1$.  Each type therefore owns the same $k$ stage positions in every partition
block, so one template and pipeline schedule serve the entire model even
though the stages' operator contents and execution costs may differ.  A replica of type~$m$ has resident batch size~$b_m$, divided into $\mu$
microbatches of size $\widehat{b}_m=b_m/\mu$.

\paragraph{Communication-aware latency model.}
Plan latency depends on both stage execution and communication across stage
boundaries.  For each stage~$j$, let
$t_j=f_{\mathrm{lat}}(\cdot)$ denote its profiled execution latency under the
assigned device-group configuration and microbatch size.  Let
$u_j=f_{\mathrm{comm}}(\cdot)$ denote the profiled transfer latency after
stage~$j$, including any redistribution required by the source and
destination parallel layouts.  The boundary after stage~$S$
connects to stage~1 of the next partition block and is omitted after the final
block.  With $B_{\mathrm{total}}$ repeated partition blocks, the steady-state
pipeline round-trip time is
$T_{\mathrm{rtt}} = f_{\mathrm{rtt}}(\mu,B_{\mathrm{total}},\mathbf{t},\mathbf{u})$,
where $\mathbf{t}=(t_1,\ldots,t_S)$ and
$\mathbf{u}=(u_1,\ldots,u_S)$.
We evaluate $f_{\mathrm{rtt}}$ by schedule simulation, which enforces
execution and transfer dependencies, models resource contention, and permits
computation--communication overlap only when the schedule and resources
allow.  Appendices~\ref{sec:placement}
and~\ref{sec:scheduling} provide the complete placement, batching, execution,
and communication definitions.

\paragraph{Design-space expressiveness.}
The planner can isolate expensive
full-attention operators, combine efficient-attention operators with neighboring
projections, or map repeated stage positions to heterogeneous device-group
types while reusing one block-level template.  This flexibility directly addresses the layerwise imbalance of
Section~\ref{sec:cad_limits}.  AFD and CAD are recovered as
restricted two-way templates: AFD separates complete attention modules from
FFN or MoE operators, whereas CAD isolates only the core attention kernels;
colocated serving is the single-stage template.

\label{sec:opt-problem}
\paragraph{Optimization problem.}
A serving plan is specified by
\[
  \mathcal{P}
  = \bigl(S,K,\mathbf{q},
  \bigl((h_m,\pi_m,b_m)\bigr)_{m=1}^{K},\mu\bigr),
\]
where $\mathbf{q}$ is the stage template defined above.  The variables $S$, $K$,
$\mu$, and $b_m$ are positive integers with $K\mid S$ and $\mu\mid b_m$ for
every type~$m$.  Each $h_m$ is drawn from the available hardware types, and
$\pi_m$ is a parallelism strategy supported by~$h_m$.  The indexed sequence
of device-group configurations is ordered because stage~$j$ is assigned to
type~$\sigma(j)$.

\begin{subequations}
\label{eq:full-opt}
\begin{align}
  \min_{\mathcal{P}}\quad
  \mathrm{CPT}_{\mathrm{plan}}
  &=
  T_{\mathrm{rtt}}\sum_{m=1}^{K}\frac{c_m}{b_m}
  \label{eq:objective} \\
  \text{s.t.}\quad
  T_{\mathrm{rtt}} &\le T_{\mathrm{SLO}}, \\
  M_{\mathrm{weights},m}
  + \mu\,\mathrm{KV}_m(\widehat{b}_m,s)
  &\le M_m,\qquad m=1,\ldots,K .
  \label{eq:memory}
\end{align}
\end{subequations}

The objective in~\eqref{eq:objective} minimizes serving cost per generated
token.  For device-group type~$m$, $c_m$ is the unit-time cost of one replica.
Thus, $\sum_m c_m/b_m$ is the aggregate
cost rate per global resident request; since each resident request produces
one token per pipeline round trip, multiplying by $T_{\mathrm{rtt}}$ gives the
steady-state cost per token.
Appendix~\ref{app:cost-derivation} derives
this objective from the global batch size and replica counts.  The constraints
require the pipeline round-trip time to meet the TPOT SLO and each replica to
fit its assigned model weights and the resident KV cache for all $\mu$
microbatches at sequence length~$s$; Appendix~\ref{sec:constraints} provides
the complete memory notation.

\paragraph{Coupled search challenge.}
The planning decisions cannot be optimized independently: stage boundaries,
device-group configurations, parallelism strategies, resident batch sizes, and
microbatch count jointly determine execution latency, transfer overhead, memory
feasibility, and replica count.  A lower-cost device may increase total cost if
its smaller feasible batch size requires more replicas, while a compute-balanced
partition may miss the TPOT SLO after boundary transfers are scheduled.
Section~\ref{sec:algorithms} exploits partition-block regularity and Pareto
structure to search this coupled space efficiently.

\section{Algorithms}
\label{sec:algorithms}

\makeatletter
\providecommand*{\theHALG@line}{\thealgorithm.\arabic{ALG@line}}
\renewcommand*{\theHALG@line}{\thealgorithm.\arabic{ALG@line}}
\makeatother

The serving-plan search in \cref{sec:opt-problem} is difficult for three
reasons.  First, independently placing every legal stage boundary
creates a combinatorial partition space.  Second, each device-group type,
which we call an \emph{owner} because it owns fixed stage positions in every
partition block, admits many hardware, parallelism, and batching
configurations.  Third, the cross-owner product of these configurations is too large to
enumerate and simulate exhaustively.  Our planner addresses
them in order: it parameterizes partitions with
regularity-aware sub-block templates, constructs a network-safe Pareto frontier
for each owner, and combines the surviving frontiers using exact
branch-and-bound.  Only combinations that survive all three reductions reach
the schedule simulator.

\subsection{Regularity-Aware Sub-Block Templates}
\label{sec:algo-subblock}

An unrestricted partition that chooses every stage boundary independently
produces $\binom{N_{\mathrm{block}}-1}{S-1}$ templates for a block with
$N_{\mathrm{block}}$ operators and $S$ stages, even though the same operator
order often repeats across many layers.  This redundancy is especially large
in hybrid-attention models: layers may use different attention variants while
retaining the same sequence of attention projections, attention core, and
FFN or MoE operators.  We remove it by searching a smaller
\emph{sub-partition block}.  Let $L_{\mathrm{block}}$ be the number of layers
in a partition block, and let $L_{\mathrm{sub}}$ be a sub-block length in
layers that divides $L_{\mathrm{block}}$ and preserves the repeated layer
pattern; each block then contains $n=L_{\mathrm{block}}/L_{\mathrm{sub}}$
sub-block repetitions.  For a chosen number of owners~$K$, the planner selects
a legal contiguous template $\mathbf{q}_{\mathrm{sub}}=(q_0,\ldots,q_K)$
inside one sub-block and tiles it over the $n$ repetitions.  Stage
position~$m$ in every repetition is assigned to owner~$m$, giving $S=nK$
stages per partition block (Section~\ref{sec:formulation}'s $S=kK$ with
$k=n$).  This parameterization reduces the number of
templates to $\binom{N_{\mathrm{block}}/n-1}{K-1}$ while retaining
layer-dependent execution costs after the template is expanded; it is a
structural restriction of the searched plan family, not a cost-based pruning
heuristic.

\begin{figure}[!t]
    \centering
    \includegraphics[width=\columnwidth]{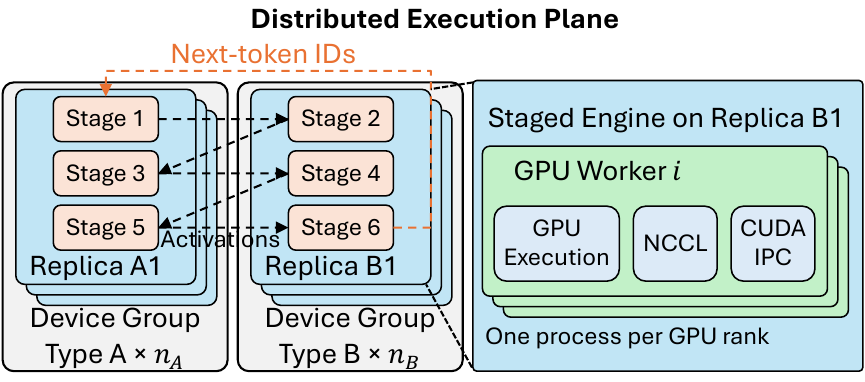}
    \caption{Distributed execution plane. Black dashed arrows carry
    activations between stages; the orange dashed path returns next-token IDs
    to the first-stage workers.}
    \label{fig:system-overview}
\end{figure}

\begin{algorithm}[!t]
\caption{Regularity-Aware Exact Serving-Plan Search}
\label{alg:serving-plan-search}
\small
\begin{algorithmic}[1]
\Require Model operator pattern, finite search grid, profiles, TPOT SLO
\Ensure Cheapest feasible plan in the structured space
\State $\mathrm{best}\leftarrow\infty$
\For{each regular sub-block and legal template $\mathbf{q}_{\mathrm{sub}}$}
  \State Tile $\mathbf{q}_{\mathrm{sub}}$ over the model
  \For{each feasible $(B,\mu)$}
    \State Build network-safe frontiers $\mathcal{F}_1,\ldots,\mathcal{F}_K$ from memory-feasible owner points
    \State \Call{Explore}{$1,\emptyset$}
  \EndFor
\EndFor
\State \Return best plan
\Procedure{Explore}{$i,P$}
  \State \textbf{if} optimistic resource and cost bounds for $P$ rule out a feasible improvement over $\mathrm{best}$ \textbf{then} \Return
  \If{$i>K$}
    \State Tighten bounds; if $P$ survives, simulate and update $\mathrm{best}$ if feasible and improved
    \State \Return
  \EndIf
  \For{each $p\in\mathcal{F}_i$}
    \State \Call{Explore}{$i+1,P\cup\{p\}$}
  \EndFor
\EndProcedure
\end{algorithmic}
\end{algorithm}

\subsection{Network-Safe Per-Owner Pareto Frontiers}
\label{sec:network-safe-frontier}

The structural reduction still leaves many configurations for every owner.
Given a template, global resident batch~$B$, and microbatch count~$\mu$, a
feasible point~$p$ for owner~$m$ chooses a hardware and parallelism
configuration and a per-replica microbatch size~$\widehat{b}_m(p)$, which fix
an integral replica count.  Profiling gives the execution
latencies $\mathbf{T}_m(p)=\bigl(t_j(p)\bigr)_{j\in\mathcal{S}_m}$ of the
stage positions $\mathcal{S}_m=\{j:\sigma(j)=m\}$ owned by~$m$.  The point also has normalized cost
and GPU coordinates $r_m(p)=c_m(p)/b_m(p)$ and $\rho_m(p)=g_m(p)/b_m(p)$,
where $b_m(p)=\mu\widehat{b}_m(p)$ and $g_m(p)$ is the GPU count per replica.

Latency and cost alone are insufficient for safe local pruning.  Two points
with similar execution latency can induce different replica counts, total GPU
use, and per-endpoint transfer sizes when combined with neighboring owners.  We
therefore compare points using
\begin{equation}
\label{eq:frontier-coordinates}
  \mathbf{z}_m(p)
  =\bigl(\mathbf{T}_m(p),r_m(p),\rho_m(p),
          \widehat{b}_m(p)\bigr).
\end{equation}
Point~$p$ dominates point~$p'$ only if both use the same tensor-parallel
degree, $p$ is no worse in every coordinate of
\eqref{eq:frontier-coordinates}, and it is strictly better in at least one;
the degree restriction is needed because transfer volume varies
non-monotonically with tensor parallelism.
The GPU coordinate preserves the planner's preference for fewer GPUs among
otherwise equal plans; the microbatch coordinate preserves the inputs of the
batch-dependent transfer model.
This dominance relation yields the network-safe frontier~$\mathcal{F}_m$.
Retaining $\mathbf{T}_m$ as a per-occurrence vector prevents a point that
accelerates one layer type but slows another from being removed, so the
frontier eliminates only configurations that cannot improve any feasible
complete plan.

\subsection{Exact Branch-and-Bound over Frontier Products}
\label{sec:frontier-branch-bound}

Explicitly evaluating $\mathcal{F}_1\times\cdots\times\mathcal{F}_K$ can
still require exponentially many schedule simulations, so the planner explores
the product with depth-first branch-and-bound.  For each prefix of assigned
owners, it forms an optimistic completion from the best remaining coordinates
of the unassigned frontiers: mandatory serialized compute work lower-bounds
each owner's latency contribution, and mandatory transfer work at each network
endpoint lower-bounds communication, with unassigned endpoints minimized
independently over feasible replica counts, keeping the estimate optimistic.
The larger of the two is a valid lower bound on $T_{\mathrm{rtt}}$, and
multiplying it by the suffix-minimal normalized cost lower-bounds the
objective.  A subtree is discarded if its optimistic completion violates the
TPOT SLO or cannot improve the incumbent.  Once every owner is assigned, the
bounds are recomputed with the complete configuration, and only survivors
invoke the schedule simulator.  Appendix~\ref{app:bnb-bounds} states these
bounds formally.

The schedule simulator remains authoritative for every surviving candidate,
so the planner is exact over the configured finite hardware, batching, and
template grid; its worst-case complexity remains $\prod_m|\mathcal{F}_m|$ when no
bound prunes.

\section{System Design and Implementation}
\label{sec:implementation}

\sys is implemented in 32K lines of Python and C++ on the vLLM GPU model
runner~\citep{kwon2023pagedattention}; each worker loads only its assigned
model components and state. Inter-node activation transfers use point-to-point
NCCL, and intra-node fanout uses CUDA interprocess communication (IPC) peer
copies with GPU doorbells. The planner-selected batch
size fixes the plan's scheduling capacity: logical slots, microbatch shapes,
stage placement, and per-rank action order are static, so CUDA graphs and
buffers are bound once, while token, position, and KV-cache metadata change
between iterations.

\subsection{System Overview and Execution Flow}
\label{sec:impl-overview}

\paragraph{Runtime organization.}
Figure~\ref{fig:system-overview} shows the distributed execution plane after
plan installation. A \sys deployment consists of a global coordinator (outside
the figure) and replicas of the device-group types selected by the plan; each
replica runs a staged engine with one GPU worker process per GPU rank. The
coordinator acts only at setup, instantiating the replicas and installing the
plan; during decoding, workers execute the compiled stage schedule without
coordinator or RPC intervention between stages and microbatches.

\begin{table}[!t]
    \centering
    \papertablestyle
    \captionsetup{font={normalsize,bf}}
    \caption{GPU specifications~\citep{nvidia_h100_specs,nvidia_l40s_specs,nvidia_a100_specs}.
    Prices are averaged across selected public clouds.}
    \label{tab:gpu-specifications}
    \setlength{\tabcolsep}{3pt}
    \begin{tabular}{@{}l|c|c|c|c@{}}
        \toprule
        \multicolumn{1}{@{}c|}{\textbf{GPU type}} &
        \makecell{\textbf{Memory BW}\\\textbf{(GB/s)}} &
        \makecell{\textbf{Memory}\\\textbf{(GB)}} &
        \makecell{\textbf{BF16}\\\textbf{(TFLOPS)}} &
        \makecell{\textbf{Price}\\\textbf{(\$/GPU-h)}} \\
        \midrule
        H100 SXM & 3,350 & 80 & 989    & 3.49 \\
        L40S     &   864 & 48 & 362.05 & 1.09 \\
        A100 SXM & 2,039 & 80 & 312    & 1.74 \\
        \bottomrule
    \end{tabular}
\end{table}

\paragraph{Request and stage execution.}
First-stage workers consume the current token IDs of batched requests. Each
microbatch traverses the ordered stages, transferring activations whenever
consecutive stages reside on different replicas; terminal workers produce
next-token IDs, which return to the first-stage workers on the GPU
(Section~\ref{sec:impl-optimizations}). A stage runs once its input activation
has arrived, enforced by receive-completion events, and the preceding compute
action in its worker's fixed action order has completed; independent
computation and communication still overlap.

\subsection{Optimizations}
\label{sec:impl-optimizations}

Fine-grained disaggregation creates many short compute and transfer actions per
token. If each traverses Python or becomes a separate network operation, its
overhead can erase the benefit of finer placement. We apply three optimizations.

\paragraph{Single native submission.}
A naive executor returns to Python for every stage, send, and receive, placing
host dispatch between otherwise short GPU actions. Instead, \sys compiles the
complete per-rank sequence; for each token, one native submission enqueues all
CUDA-graph launches, staging copies, communication rounds, and event
dependencies, eliminating Python dispatch, allocation, and RPCs between stages
and microbatches.

\paragraph{GPU-resident decoding.}
A centralized implementation would copy generated token IDs to the host,
assemble and redistribute them, and rebuild execution metadata at every step,
serializing a CPU--GPU round trip into each token. In \sys, terminal workers compute token
IDs on the GPU, one NCCL all-reduce distributes the token vector to embedding
owners, and tokens stay device-resident between iterations; only global rank
zero retains the token history for eventual host-side output. Each stage
reuses preconstructed metadata and preallocated buffers, updating only the
tensors that change between iterations.

\begingroup
\emergencystretch=1em
\paragraph{Low-overhead communication.}
Our runtime packs compatible tensors into shared buffers and batches transfers
into NCCL rounds, reducing per-operation overhead without increasing
activation payload. When an unsharded activation crosses a stage boundary into
a TP group, the runtime sends one inter-node copy to a destination leader
rather than one copy per rank; the leader fans it out to the remaining TP
ranks locally through CUDA IPC, and leader assignment rotates across routes to
balance cross-node traffic.
\par\endgroup

\begin{table}[!t]
    \centering
    \papertablestyle
    \captionsetup{font={normalsize,bf}}
    \caption{Evaluated models and TPOT SLOs. Eff. Attn. is the efficient
    attention mechanism interleaved with full attention; Pattern is the ratio
    of efficient- to full-attention layers.}
    \label{tab:evaluated-models}
    \setlength{\tabcolsep}{2.5pt}
    \begin{tabular}{@{}l|c|c|c|c|c@{}}
        \toprule
        \multicolumn{1}{@{}c|}{\textbf{Model}} & \textbf{MoE} &
        \makecell{\textbf{Eff.}\\\textbf{Attn.}} & \textbf{Pattern} &
        \makecell{\textbf{Strict}\\\textbf{SLO (ms)}} &
        \makecell{\textbf{Relaxed}\\\textbf{SLO (ms)}} \\
        \midrule
        Gemma-3-27B        & \tableno  & SWA    & 5:1 & 25 & 60 \\
        Qwen3-Next-80B-A3B & \tableyes & Linear & 3:1 & 40 & 80 \\
        \bottomrule
    \end{tabular}
\end{table}

\begin{figure*}[t]
    \centering
    \includegraphics[width=\textwidth]{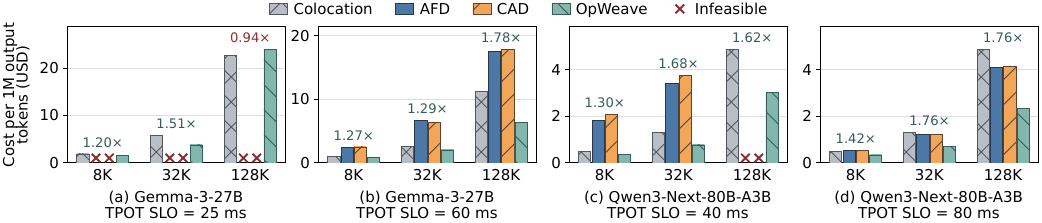}
    \Description{Four framed grouped bar charts in one horizontal row show
    Gemma-3-27B under 25- and 60-millisecond TPOT SLO labels, followed by
    Qwen3-Next-80B-A3B under 40- and 80-millisecond labels. Each panel has
    8K, 32K, and 128K context groups with four slots ordered as Colocation,
    AFD, CAD, and \sys{}. Bar heights retain measured USD per million output
    tokens, with an independent zero-based linear scale for each panel.
    Crosses mark the eight infeasible configurations. Small labels above
    each context group show the cheapest feasible Colocation, AFD, or CAD
    baseline cost divided by \sys{} cost; a ratio below one indicates higher
    \sys{} cost. Panel captions are below the charts
    and a shared legend is above.}
    \caption{Measured serving cost across context lengths and TPOT SLOs.
    Crosses mark infeasible configurations.
    Labels above each group show $g\times$ cheaper, where $g$ is the best
    feasible baseline cost divided by \sys{} cost.}
    \label{fig:homogeneous-measured-cost}
\end{figure*}

\begin{figure*}[t]
    \centering
    \includegraphics[width=\textwidth]{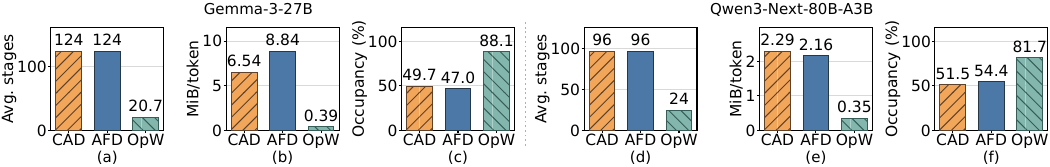}
    \Description{Six panels in one horizontal row, with Gemma-3-27B in the
    first three and Qwen3-Next-80B-A3B in the last three. Panels a and d show
    average full-model pipeline stage counts for CAD, AFD, and \sys{}:
    124, 124, and 20.67 for Gemma; 96, 96, and 24 for Qwen3-Next.
    Panels b and e show mean inter-node stage-to-stage payload in MiB per
    output token: 6.54, 8.84, and 0.39 for Gemma; 2.29, 2.16, and 0.35 for
    Qwen3-Next. Panels c and f show price-weighted compute occupancy:
    49.71, 47.01, and 88.08 percent for Gemma; 51.45, 54.37, and 81.74 percent
    for Qwen3-Next.}
    \caption{Homogeneous-machine evaluation results. OpW denotes \sys{}.
    (a) and (d): Mean number of pipeline stages.
    (b) and (e): Mean inter-node stage-to-stage payload per output token (MiB).
    (c) and (f): Mean price-weighted compute occupancy (Section~\ref{sec:e2e-homogeneous}).}
    \label{fig:homogeneous-evaluation}
\end{figure*}

\begin{figure}[t]
    \centering
    \captionsetup[subfloat]{font=paperplot,labelfont={},textfont={}}
    \subfloat[Gemma-3-27B.]{%
        \scalebox{1}[0.9]{\includegraphics[width=0.48\linewidth]{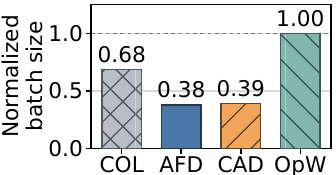}}%
        \label{fig:normalized-batch-size-gemma}%
    }
    \hfill
    \subfloat[Qwen3-Next-80B-A3B.]{%
        \scalebox{1}[0.9]{\includegraphics[width=0.48\linewidth]{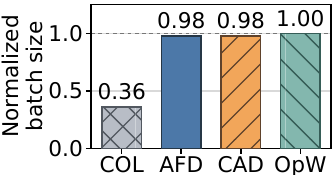}}%
        \label{fig:normalized-batch-size-qwen}%
    }
    \Description{Two bar charts in one horizontal row show mean normalized
    per-GPU batch size at the relaxed TPOT SLO for Colocation, AFD, CAD, and
    \sys{}. The Gemma-3-27B values are 0.68, 0.38, 0.39, and 1.00. The
    Qwen3-Next-80B-A3B values are 0.36, 0.98, 0.98, and 1.00.}
    \caption{Mean per-GPU batch size at the relaxed TPOT SLO, normalized to
    \sys{} within each context and averaged across contexts. Higher is better.}
    \label{fig:normalized-batch-size}
\end{figure}

\begin{figure}[!t]
    \centering
    \includegraphics[width=\linewidth]{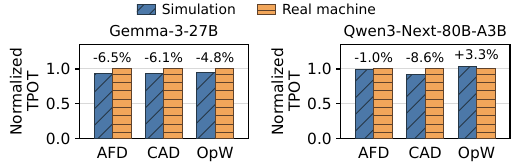}
    \caption{Simulated and measured TPOT for AFD, CAD, and \sys{}, normalized
    by measured TPOT. Labels show signed simulation errors.}
    \Description{Two grouped bar charts in one horizontal row compare
    simulation and real-machine TPOT on the same zero-based normalized
    scale, with Gemma-3-27B on the left and Qwen3-Next-80B-A3B on the right,
    and one shared legend centered above the panels.
    Each pair is normalized by its real-machine TPOT, so all real-machine
    bars equal 1. For AFD, CAD, and \sys{} respectively, Gemma simulation bars
    are 0.935, 0.939, and 0.952; Qwen3-Next simulation bars are 0.990,
    0.914, and 1.033. Labels above the groups show signed percentage
    differences: -6.5\%, -6.1\%, and -4.8\% for Gemma, and -1.0\%,
    -8.6\%, and +3.3\% for Qwen3-Next.}
    \label{fig:simulator-fidelity}
\end{figure}

\section{Evaluation}
\label{sec:evaluation}

\subsection{Evaluation Setup}
\label{sec:eval-setup}

\paragraph{Devices and Models.} Table~\ref{tab:gpu-specifications} lists the
GPUs used in our evaluation. We conduct the homogeneous evaluation on a
cluster of H100 nodes connected by RDMA over Converged Ethernet (RoCE). For
the simulator fidelity tests, we use H100 and L40S GPUs on AWS, where RDMA
transfers stage GPU data through host memory.
We evaluate two representative hybrid-attention models~\citep{gemma3,qwen3_next},
with their architectures and TPOT SLOs summarized in
Table~\ref{tab:evaluated-models}.

\paragraph{Metrics.} We use serving cost per million output tokens as our
primary efficiency metric and time per output token (TPOT) as our latency
metric. A configuration satisfies a target TPOT service-level objective (SLO)
if its p95 step-level TPOT across the measured iterations does not exceed the
target; for each serving policy, we report the lowest-cost configuration that
satisfies the SLO.

\paragraph{Baselines.} We compare \sys against three baselines: colocated
serving (COL), attention--FFN disaggregation
(AFD)~\citep{zhu2025megascaleinfer,stepfun2025step3}, and core-attention
disaggregation (CAD). COL is implemented with vLLM
0.26.0~\citep{kwon2023pagedattention} and executes all operators on GPUs
within the same node. AFD and CAD are implemented in our \sys runtime
as fixed-partition special cases. AFD separates complete
attention modules from FFN/MoE operators, whereas CAD isolates core-attention
kernels from all remaining operators; their two device groups may be placed on
different nodes. This setup gives COL a well-optimized production implementation and keeps
the runtime identical across disaggregated policies.

\paragraph{Search Configuration.} For each model, workload, and TPOT SLO, our
planner searches the serving-plan space of Section~\ref{sec:algorithms} for
the lowest-cost plan meeting the SLO. We restrict each plan to at most 32
GPUs, three device-group types, four replicas per type, and four microbatches.
\sys and all baselines use the same search configuration and resource
constraints; their admissible plans differ only in the imposed disaggregation
policy. The planner returns the optimal feasible plan within this configured
finite search space.

\paragraph{Workloads.} We evaluate steady-state decoding using long-context
workloads derived from OpenThoughts3-1.2M~\citep{guha2025openthoughts}. We
tokenize the dataset to obtain an empirical distribution of generation-time
context lengths and rescale it to construct synthetic request cohorts with
mean context lengths of 8K, 32K, and 128K tokens. Requests are constructed
deterministically, and all serving policies use the same distribution and
construction procedure, ensuring comparable workloads.

\begin{figure*}[!t]
    \centering
    \includegraphics[width=\textwidth]{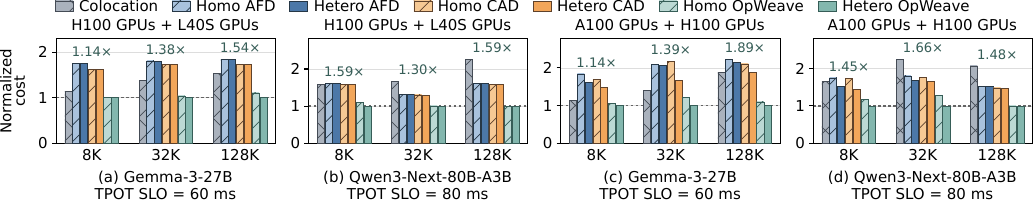}
    \caption{Simulated serving cost across context lengths and TPOT SLOs,
    normalized to heterogeneous \sys{} within each context.
    Labels show $g\times$ cheaper, where $g$ is the best feasible non-\sys{}
    baseline cost divided by heterogeneous \sys{} cost.}
    \label{fig:heterogeneous-cost}
\end{figure*}

\subsection{Homogeneous End-to-End Evaluation}
\label{sec:e2e-homogeneous}

We evaluate \sys{} and all baselines on H100 GPUs across both models, both SLOs
(Table~\ref{tab:evaluated-models}), and all three context lengths.
As shown in Figure~\ref{fig:homogeneous-measured-cost}, \sys{} achieves the lowest serving cost in
11 of the 12 evaluated model--context--SLO settings. \sys{} is up to $1.78\times$ cheaper than the best feasible baseline for
Gemma-3-27B and up to $1.76\times$ cheaper for Qwen3-Next-80B-A3B.

\sys{} finds feasible plans in all evaluated scenarios and achieves lower serving
costs than AFD and CAD wherever they are feasible. \emph{Its first advantage
is lower latency through fewer pipeline stages.} Whereas AFD and CAD impose
two stages per layer, \sys{} selects stage boundaries across an entire partition
block, enabling a larger, more flexible partition space. As shown in
Figure~\ref{fig:homogeneous-evaluation}(a) and (d), the selected \sys{} plans average
only 20.7 stages for Gemma-3-27B and 24 for Qwen3-Next-80B-A3B, compared with
124 and 96, respectively, for both baselines. Fewer stage boundaries reduce
network transfers: Figure~\ref{fig:homogeneous-evaluation}(b) and (e) show that \sys{}
reduces mean inter-node data transferred per output token by factors of 16.9
and 22.8 relative to CAD and AFD, respectively, for Gemma, and by factors of
6.5 and 6.1 for Qwen3-Next. Because inter-stage communication lies on the
execution critical path of each microbatch, this reduction lowers decoding
latency, helping \sys{} satisfy strict TPOT SLOs in settings where neither
baseline finds a feasible plan.

\emph{\sys{}'s second advantage is higher hardware occupancy by model execution,
reducing idle time during decoding.} We report system-level occupancy as the
average of per-replica occupancy, weighting each replica type by its price
times its replica count, i.e., the fraction of hardware spending that performs
model execution. As shown in
Figure~\ref{fig:homogeneous-evaluation}(c) and (f), \sys{} achieves mean compute
occupancy above 80\% for both models, while CAD and AFD leave roughly half of
the hardware time unused. One reason for the baselines' lower occupancy is
heterogeneity across layers: the selected hybrid models interleave attention
mechanisms with substantially different execution latencies, making their
fixed partitions difficult to balance. For example, at 128K context and batch
size 1 on one H100, our profiles estimate 700~$\mu$s for a Qwen3-Next
full-attention module versus 58~$\mu$s for linear attention, a $12\times$
gap. \sys{}'s flexible partition
boundaries allow it to balance work across these layers. Moreover, AFD and CAD's high
network transfer volume leaves limited latency headroom for increasing the
number of microbatches to overlap work across device groups. Even under
Gemma's relaxed 60-ms SLO, their selected plans still use only one microbatch.
The two device groups therefore execute dependent stages sequentially, leaving
one group idle while the other works.

\emph{\sys{} also alleviates the memory contention that limits batching in
colocated serving.} The attention KV cache competes with model weights for
GPU memory, so longer contexts reduce the maximum feasible batch size. \sys{}
mitigates this constraint by decoupling GEMM batching from KV-cache memory
pressure. Consistent with our theoretical analysis,
Figure~\ref{fig:normalized-batch-size} shows that \sys{} achieves larger mean
per-GPU batch sizes than colocation for both models under the relaxed SLOs.
AFD and CAD similarly alleviate this constraint for Qwen3-Next, achieving
batch sizes close to \sys{}'s. For Gemma, however, their high communication
latency forces them to use smaller batches to satisfy the TPOT SLO, resulting
in lower per-GPU batch sizes even than colocation.

\subsection{Simulator Fidelity}
\label{sec:simulator-fidelity}

We use the simulator for large-scale heterogeneous evaluation because
real-machine experiments at this scale are prohibitively expensive and
heterogeneous clusters with high-speed inter-node connections are difficult
to allocate on public clouds. We validate it using two H100 and two L40S GPUs
for Gemma-3-27B and four of each for Qwen3-Next-80B-A3B. As shown in
Figure~\ref{fig:simulator-fidelity}, the average absolute simulation errors
across AFD, CAD, and \sys{} are 5.8\% for Gemma-3-27B and 4.3\% for
Qwen3-Next-80B-A3B.

\subsection{Heterogeneous End-to-End Evaluation}
\label{sec:e2e-heterogeneous}

We use our simulator to evaluate \sys{} and all baselines on two heterogeneous
setups, H100 with L40S and H100 with A100
(Table~\ref{tab:gpu-specifications}), using the relaxed TPOT SLOs and all
three context lengths.
As shown in Figure~\ref{fig:heterogeneous-cost}, heterogeneous \sys{} achieves
lower serving cost than all non-\sys{} baselines in all 12 evaluated
model--hardware--context settings. On H100 with L40S, heterogeneous \sys{}
is up to $1.54\times$ cheaper for Gemma-3-27B and $1.59\times$ cheaper for
Qwen3-Next-80B-A3B than the best feasible non-\sys{} baseline.
On H100 with A100, heterogeneous \sys{} is up to $1.89\times$ and $1.66\times$
cheaper, respectively. \sys{} thus retains its cost advantage over non-\sys{} baselines in
heterogeneous settings.

Compared with homogeneous \sys{}, heterogeneous \sys{} is $1.04\times$ cheaper on
average (arithmetic mean of per-setting cost ratios across both models and
contexts) and up to $1.11\times$ cheaper on H100 with L40S. On H100 with A100, heterogeneous \sys{} is $1.14\times$
cheaper on average and up to $1.28\times$ cheaper. The larger benefit on H100 with A100 is
consistent with the analysis in Section~\ref{sec:hetero_vs_homo}: a greater
disparity in the GPUs' compute-to-memory-bandwidth ratios permits a higher
upper bound on the idealized gain from heterogeneous assignment.
Using the specifications in Table~\ref{tab:gpu-specifications}, the hardware
ratio $\eta$ is 1.42 for H100 with L40S and 1.93 for H100 with A100,
yielding idealized gain bounds of $1.10\times$ and $1.19\times$, respectively,
under Eq.~\ref{eq:hetero_bound}. The measured gains can exceed these idealized bounds because the bounds
assume both deployments fully disaggregate memory-bound attention from
compute-bound GEMM-based operators, which neither deployment achieves in
practice.

\begin{figure}[!t]
    \centering
    \includegraphics[width=\linewidth]{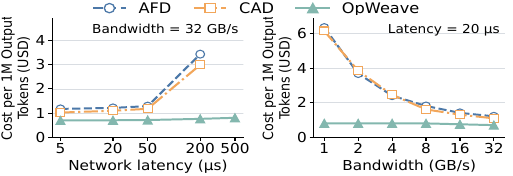}
    \caption{Network sensitivity of \sys{}, AFD and CAD.}
    \Description{Two line plots compare AFD, CAD, and \sys{}. Latency is
    swept over 5, 20, 50, 200, and 500 microseconds at 32 GB/s; bandwidth
    is swept over 1, 2, 4, 8, 16, and 32 GB/s at 20 microseconds.
    \sys{} has the lowest reported cost at every comparable setting.
    AFD and CAD have no reported winner at 500 microseconds.}
    \label{fig:network-sensitivity}
\end{figure}

\begin{table}[!t]
    \centering
    \papertablestyle
    \captionsetup{font={normalsize,bf}}
    \setlength{\tabcolsep}{1.5pt}
    \begin{tabular}{@{}c|c|c|c|c@{}}
        \toprule
        \textbf{HW} & \textbf{Method} & \textbf{Candidates} &
        \textbf{Simulations} & \textbf{Time (s)} \\
        \midrule
        \multirow{4}{*}{H100}
          & Full     & 4.78M (1.00$\times$)   & 4.78M (1.00$\times$)  & 228.3 (1.00$\times$) \\
          & No-PF    & 8.83M (1.85$\times$)   & 8.83M (1.85$\times$)  & 264.4 (1.16$\times$) \\
          & No-B\&B  & 219.79M (45.96$\times$) & 4.78M (1.00$\times$)  & 290.3 (1.27$\times$) \\
          & Neither  & 293.24M (61.31$\times$) & 8.83M (1.85$\times$)  & 330.5 (1.45$\times$) \\
        \midrule
        \multirow{4}{*}{\makecell[l]{A100\\+H100}}
          & Full     & 13.89M (1.00$\times$)    & 13.89M (1.00$\times$) & 446.7 (1.00$\times$) \\
          & No-PF    & 52.14M (3.75$\times$)    & 52.14M (3.75$\times$) & 776.4 (1.74$\times$) \\
          & No-B\&B  & 630.97M (45.42$\times$)   & 13.89M (1.00$\times$) & 636.6 (1.42$\times$) \\
          & Neither  & 2.33B (167.79$\times$) & 52.14M (3.75$\times$) & 1,495.8 (3.35$\times$) \\
        \bottomrule
    \end{tabular}
    \caption{Planner search ablation. \emph{Full} enables Pareto-frontier (PF)
    pruning and branch-and-bound (B\&B); \emph{No-PF}, \emph{No-B\&B}, and
    \emph{Neither} disable the corresponding optimizations. Parentheses show
    degradation relative to \emph{Full} on the same hardware. Lower is better;
    all methods find the same best plan.}
    \label{tab:planner-search-ablation}
\end{table}

\subsection{Ablation Study and Sensitivity Study}
\label{sec:eval-ablation}

\paragraph{Planner Search Ablation.}
This ablation quantifies the benefits of Pareto-frontier pruning and
branch-and-bound for planner search. We search for Gemma-3-27B plans at 32K
context under a 60-ms TPOT SLO, using 64 CPU cores for every run.
Table~\ref{tab:planner-search-ablation} reports candidates considered,
simulator invocations, and end-to-end search time. Because inexpensive memory
and SLO checks discard many candidates before the costly simulator is invoked,
reducing candidate count does not translate proportionally into wall-clock
time. Pareto-frontier pruning reduces simulator invocations, whereas
branch-and-bound removes candidates before simulation. Disabling PF increases
simulator invocations by $1.85\times$ on H100 and $3.75\times$ on A100+H100,
increasing wall-clock time by $1.16\times$ and $1.74\times$, respectively.
Disabling B\&B instead increases considered candidates by $45.96\times$ and
$45.42\times$ without increasing simulator invocations; these are rejected by
the cheaper checks, so wall-clock time increases by only $1.27\times$ and
$1.42\times$. Together, they make search $1.45\times$ faster on H100 and
$3.35\times$ on A100+H100 than using neither.

\paragraph{Network Sensitivity.}
We use simulation to evaluate the sensitivity of \sys{}, AFD, and CAD to network
conditions, varying inter-node transfer latency and bandwidth for Gemma-3-27B
at 8K context under a 60-ms TPOT SLO. Each method is reoptimized at every
setting over H100, L40S, and mixed deployments. As shown in
Figure~\ref{fig:network-sensitivity}, \sys{} is less sensitive than AFD and CAD
to both network latency and bandwidth. This robustness follows from \sys{}'s much
lower network transfer volume
(Figure~\ref{fig:homogeneous-evaluation}(b) and (e)): the remaining transfers
can be overlapped with model execution, leaving serving performance less
affected by network conditions.

\section{Related Work}
\label{sec:related_work}

\paragraph{LLM Serving Optimization.}
General-purpose LLM serving systems improve efficiency through continuous batching~\citep{yu2022orca}, paged KV-cache management~\citep{kwon2023pagedattention}, KV reuse and prefix caching~\citep{zheng2023sglang,qin2024mooncake}, prefill--decode disaggregation and scheduling~\citep{patel2024splitwise,zhong2024distserve,agrawal2024sarathi}, and speculative decoding~\citep{chen2023speculative,cai2024medusa,chen2025slosserve,chen2024sequoia,huang2025adaspec,leviathan2022speculative,li2024eagle2,li2024eagle,li2025eagle3,liu2024turbospec,miao2024specinfer,oliaro2024suffixdecoding,li2026adaserve}. These techniques substantially improve LLM serving, but they do not directly address the different bottlenecks of attention and GEMMs.

\paragraph{Operator-Level Disaggregated Serving.}
MegaScale-Infer and Step-3 AFD are closest to our setting: both use a fixed two-way split that disaggregates attention from FFN, MLP, or MoE during decoding~\citep{zhu2025megascaleinfer,stepfun2025step3}. FastDecode similarly separates attention and its KV cache from the dense layers, but offloads attention to distributed CPU workers rather than another GPU group~\citep{he2024fastdecode}. Infinite-LLM distributes attention computation and KV-cache capacity across instances, rather than general operator-class placement~\citep{lin2024infinite}. NanoFlow also operates at operation granularity, but overlaps compute, memory, and network work within a device instead of separating attention and FFN onto distinct device groups~\citep{zhu2024nanoflow}. Unlike these systems, we provide a theoretical analysis of the achievable ODS cost gains, formalize the broader serving-plan space, and show why fixed two-way splits can fail for hybrid-attention architectures such as Gemma~3, GPT-OSS, and Qwen3-Next~\citep{gemma3,gpt_oss,qwen3_next}.

\paragraph{Heterogeneous LLM Serving.}
A growing line of work serves LLMs across heterogeneous GPU resources. Helix formulates heterogeneous serving as a max-flow problem, jointly optimizing layer placement and request routing across GPU types~\citep{mei2024helix}. Other systems apply asymmetric pipeline parallelism to decentralized heterogeneous or volunteer GPUs~\citep{jiang2024hexgen,borzunov2022petals,borzunov2023distributed,tong2025parallax,wu2025deserve,li2024tpillm,kim2025flexllm,jiang2025hexgen2,jiang2025thunderserve,peng2025hexgenflow}, or exploit price-performance gaps across GPU types by routing requests to the best-suited hardware~\citep{griggs2024melange,jiang2026boute,mei2026coral}. All of these systems partition at the transformer-layer boundary or coarser; in contrast, this paper disaggregates at the operator level, matching heterogeneous GPU types to operator classes with different resource bottlenecks inside a single layer.

\section{Conclusion}
\label{sec:conclusion}

This paper presented \sys, an end-to-end framework for heterogeneous operator-level disaggregated serving that connects theory, planning, and execution. \sys develops an analytical cost model that characterizes and bounds the gains of homogeneous and heterogeneous ODS over colocated serving, a regularity-aware planner that jointly optimizes operator partitioning, hardware assignment, parallelism, and batching over the partition block abstraction, and a vLLM-based distributed runtime that executes the synthesized plans across heterogeneous device groups. In our evaluation, \sys is up to $1.78\times$ cheaper than the best feasible baseline on homogeneous GPUs and up to $1.89\times$ cheaper on heterogeneous clusters, while attaining latency SLOs. We hope the analysis and abstractions in \sys provide a foundation for future work on operator-level disaggregation across increasingly heterogeneous models and hardware.

\section*{Acknowledgments}
We used ChatGPT, Claude, and Gemini to improve the clarity and readability of the manuscript.
We also used Claude and GPT to assist with code development for this work.
This work was partially supported by NSF awards CNS-2211882 and CNS-2239351, a Sloan Research Fellowship, and research awards from Amazon, Cisco, Google, Jane Street, Meta, NVIDIA, Oracle, Qualcomm, and Samsung. 

\bibliographystyle{ACM-Reference-Format}
\bibliography{reference}

\clearpage
\appendix

\section{Analytical Model and Proofs for Section~\ref{sec:theory}}
\label{app:analytical-model}

This appendix section collects the analytical model, proofs, and CAD
feasibility details used by \Secref{sec:theory}.

\subsection{Detailed analytical model}
\label{app:analytical-model-details}

We now formalize the analytical abstraction used in \Secref{sec:analytical_setup}. We model one decode step of an LLM as the sequential execution of an ordered operator sequence
\[
\mathbf{o} = (o_1, o_2, \ldots, o_{N_{\mathrm{model}}}),
\]
where each $o_i$ denotes one computational kernel invoked during the decode step. This abstraction follows the execution pattern of current serving engines, which invoke the model's kernels in a fixed topological order during each decoding iteration.

For an operator $o_i$ executed at batch size $b$, let $D_i(b)$ denote the total number of bytes loaded from memory and let $W_i(b)$ denote the total number of floating-point operations. We characterize each GPU type by four quantities: unit-time monetary cost $c$, memory bandwidth $\beta$, memory capacity $M$, and peak compute throughput $F$. Under the exclusive-occupancy assumption, the monetary cost of operator $o_i$ is
\[
\mathrm{Cost}(o_i) = T(o_i)\cdot c,
\]
where $T(o_i)$ is the execution latency of the operator. We model this latency using a roofline-style expression:
\[
T(o_i) = \max\!\left(\frac{D_i(b)}{\beta}, \frac{W_i(b)}{F}\right).
\]
The first term is the memory-load time and the second term is the compute time, so the maximum captures whether the operator is memory-bound or compute-bound on the target GPU.

The cost of a full decode step is then the sum of the costs of all operators in the sequence:
\[
\mathrm{Cost}_{\mathrm{step}} = \sum_{i=1}^{N_{\mathrm{model}}} \mathrm{Cost}(o_i).
\]
Since one decode step produces one token per request, the GPU cost per token is obtained by dividing $\mathrm{Cost}_{\mathrm{step}}$ by the batch size $b$.

Throughout the theoretical analysis, we intentionally focus on GPU-side cost. Although a disaggregated pipeline introduces activation transfers between device groups, the analysis studies the idealized steady-state regime in which these transfers are overlapped with computation through microbatching and pipelining. Under this assumption, communication does not lie on the throughput-critical path and is therefore omitted from the analytical objective.

\subsection{Execution-regime assumptions}
\label{app:execution-regimes}

This subsection formalizes the execution-regime assumptions used in \Secref{sec:analytical_setup}. Under the roofline model above, an operator is \emph{memory-bound} when
\[
\frac{D_i(b)}{\beta} \ge \frac{W_i(b)}{F},
\]
and \emph{compute-bound} when the reverse inequality holds.

We assume that decode-phase attention lies in the memory-bound regime. During decoding, each request contributes only one query token while the attention operator must load the full KV cache associated with the prior context, which keeps arithmetic intensity low and makes latency primarily determined by memory traffic. Formally, for attention operators $o_i^{\mathrm{attn}}$, we assume
\[
\frac{D_i^{\mathrm{attn}}(b)}{\beta} \ge \frac{W_i^{\mathrm{attn}}(b)}{F}.
\]

For GEMM-dominated operators, we assume that there exists a batch-size threshold $b^*$ such that these operators become compute-bound once the batch size is sufficiently large. Formally, for GEMM operators $o_i^{\mathrm{gemm}}$, we assume that for all $b \ge b^*$,
\[
\frac{W_i^{\mathrm{gemm}}(b)}{F} \ge \frac{D_i^{\mathrm{gemm}}(b)}{\beta}.
\]
The threshold $b^*$ marks the transition at which arithmetic time overtakes memory-load time, so beyond this point further execution is limited by compute throughput rather than bandwidth.

The role of disaggregation is to make this compute-bound GEMM regime attainable. In colocated serving, GEMM operators share device memory with the KV cache, so long contexts reduce the maximum feasible batch size and can keep GEMM in a memory-bound regime. Under operator-level disaggregation, the compute side no longer needs to reserve memory for the KV cache, and requests forwarded from multiple attention-side replicas can be aggregated on the GEMM side. This decoupling allows the GEMM side to sustain a larger effective batch size and thus operate in the compute-bound regime assumed in the main-text analysis.

\subsection{Derivation and interpretation of the final cost expressions}
\label{app:cost-expression-details}

This subsection derives the cost expressions used in \Secref{sec:analytical_setup} from the operator-level roofline model.

\paragraph{Operator-class aggregates.}
Let $\mathcal{O}_{\mathrm{attn}}$ denote the set of attention operators executed in one decode step and let $\mathcal{O}_{\mathrm{gemm}}$ denote the set of GEMM-dominated operators, where the latter includes the linear projections, FFN, and MoE computation. We define the aggregate quantities
\begin{align*}
D^{\mathrm{attn}}(b) &\;=\; \sum_{o_i \in \mathcal{O}_{\mathrm{attn}}} D_i(b), \\
D^{\mathrm{gemm}}(b) &\;=\; \sum_{o_i \in \mathcal{O}_{\mathrm{gemm}}} D_i(b), \\
W^{\mathrm{gemm}}(b) &\;=\; \sum_{o_i \in \mathcal{O}_{\mathrm{gemm}}} W_i(b).
\end{align*}
These aggregate terms collect, respectively, the total attention-side memory traffic, the total GEMM-side memory traffic, and the total GEMM-side floating-point work for a batch of size $b$ in one decode step.

Substituting these class aggregates into the roofline cost model of
Appendix~\ref{app:analytical-model-details} and dividing by the batch size
yields the per-token costs as follows.

\paragraph{Homogeneous disaggregated serving.}
In homogeneous ODS, attention and GEMM-dominated operators are placed on separate device groups of the same GPU type $(c,\beta,M,F)$. Under the execution-regime assumptions of \Secref{sec:analytical_setup}, decode-phase attention is memory-bound while GEMM-dominated operators are compute-bound. Their class-level costs therefore reduce to
\[
\mathrm{Cost}^{\mathrm{attn}}_{\mathrm{step}}
=
\frac{D^{\mathrm{attn}}(b)}{\beta}\,c,
\qquad
\mathrm{Cost}^{\mathrm{gemm}}_{\mathrm{step}}
=
\frac{W^{\mathrm{gemm}}(b)}{F}\,c.
\]
Dividing the sum by $b$ yields
\[
\mathrm{CPT}^{\mathrm{hom}}
=
\frac{c}{b}
\left(
\frac{D^{\mathrm{attn}}(b)}{\beta}
+
\frac{W^{\mathrm{gemm}}(b)}{F}
\right).
\]

\paragraph{Heterogeneous disaggregated serving.}
In heterogeneous ODS, the attention side and the GEMM side may run on different GPU types. If attention runs on GPU type 1 with parameters $(c_1,\beta_1,M_1,F_1)$ and GEMM-dominated operators run on GPU type 2 with parameters $(c_2,\beta_2,M_2,F_2)$, the same regime assumptions give
\[
\mathrm{Cost}^{\mathrm{attn}}_{\mathrm{step}}
=
\frac{D^{\mathrm{attn}}(b)}{\beta_1}\,c_1,
\qquad
\mathrm{Cost}^{\mathrm{gemm}}_{\mathrm{step}}
=
\frac{W^{\mathrm{gemm}}(b)}{F_2}\,c_2,
\]
and thus
\[
\mathrm{CPT}^{\mathrm{het}}_{1,2}
=
\frac{1}{b}
\left(
\frac{D^{\mathrm{attn}}(b)}{\beta_1}c_1
+
\frac{W^{\mathrm{gemm}}(b)}{F_2}c_2
\right).
\]

The cost-efficiency metrics $\alpha_j$ and $\gamma_j$ defined in Section~\ref{sec:analytical_setup} measure monetary cost per byte of memory traffic and per FLOP, respectively. Heterogeneous specialization allows attention to use the GPU type with the lower $\alpha_j$ and GEMM-dominated operators to use the type with the lower $\gamma_j$.

\paragraph{Colocated serving.}
In colocated serving, all operators share the same device memory, so the feasible batch size is constrained by both model weights and the KV cache. Let $M_{\mathrm{weights}}$ denote the memory occupied by model weights and let $m_{\mathrm{kv}}(s)$ denote the KV-cache memory required per request at average sequence length $s$. Neglecting activation memory for simplicity, the maximum feasible batch size is
\[
b_{\max}(s)
=
\left\lfloor
\frac{M-M_{\mathrm{weights}}}{m_{\mathrm{kv}}(s)}
\right\rfloor.
\]
As $s$ grows, $m_{\mathrm{kv}}(s)$ increases and the feasible batch size shrinks.

Attention remains memory-bound under colocation, so its contribution is still
\[
\frac{D^{\mathrm{attn}}(b)}{\beta}\,c.
\]
For GEMM-dominated operators, however, the smaller feasible batch size may prevent the compute side from reaching the compute-bound regime. We therefore retain the full roofline expression,
\[
\mathrm{Cost}^{\mathrm{gemm}}_{\mathrm{step}}
=
\max\!\left(
\frac{D^{\mathrm{gemm}}(b)}{\beta},
\frac{W^{\mathrm{gemm}}(b)}{F}
\right)c,
\]
which gives, for $1 \le b \le b_{\max}(s)$,
\[
\begin{aligned}
\mathrm{CPT}^{\mathrm{coloc}}(s,b)
&=
\frac{c}{b}
\Biggl(
\frac{D^{\mathrm{attn}}(b)}{\beta}
\\
&\qquad+
\max\!\left(
\frac{D^{\mathrm{gemm}}(b)}{\beta},
\frac{W^{\mathrm{gemm}}(b)}{F}
\right)
\Biggr).
\end{aligned}
\]

\paragraph{Intuition for the small-batch regime.}
The max term in the colocated expression is important because GEMM efficiency depends on batch size. When the feasible batch size remains large enough, GEMM can stay compute-bound and its contribution is governed by $W^{\mathrm{gemm}}(b)/F$. But once KV-cache growth drives the feasible batch size down, GEMM may enter a small-batch regime in which loading weights dominates the arithmetic work. In that case,
\[
\frac{D^{\mathrm{gemm}}(b)}{\beta}
\;>\;
\frac{W^{\mathrm{gemm}}(b)}{F},
\]
so GEMM becomes memory-bound and the GPU's compute capability is underutilized. This is precisely the inefficiency that operator-level disaggregation removes: by decoupling the compute side from KV-cache memory pressure, disaggregation allows GEMM-dominated operators to recover the larger effective batch sizes needed to operate in the compute-bound regime.

\paragraph{Conditional achievability via CAD}
\label{app:cad_achievability}
The idealized disaggregated cost expressions above are attainable by the
Core-Attention Disaggregation (CAD) partition under pipeline balance.
CAD assigns the core attention kernels to the attention device group and
assigns the remaining GEMM-dominated operators, including projections,
FFN, and MoE computation, to the rest-of-model device group. Under the
regime assumptions in Appendix~\ref{app:execution-regimes}, the attention
side therefore incurs bandwidth cost while the rest-of-model side incurs
compute cost. If the two stages are provisioned so that their steady-state
latencies match and inter-stage communication is overlapped by microbatching,
the realized per-token GPU cost is exactly the homogeneous or heterogeneous
ODS cost expression derived above.

\subsection{Scaling approximations and GEMM threshold}
\label{app:scaling-threshold-details}

This subsection formalizes the approximations used in \Secref{sec:analytical_setup} to obtain closed-form analytical results.

\paragraph{Attention-side memory traffic.}
During decoding, each request contributes one newly generated token, but attention must load the KV cache associated with that request's prior context. As a result, the total attention-side memory traffic grows linearly with the batch size and with the average sequence length. We therefore approximate
\[
D^{\mathrm{attn}}(b) \approx b\, m_{\mathrm{kv}}(s),
\]
where $m_{\mathrm{kv}}(s)$ denotes the KV-cache memory required per request at average sequence length $s$.

\paragraph{GEMM-side floating-point work.}
We approximate the total GEMM work in one decode step as
\[
W^{\mathrm{gemm}}(b) \approx 2bP_{\mathrm{act}},
\]
where $P_{\mathrm{act}}$ is the effective number of GEMM-side parameters activated by one request on the modeled device group after any model-parallel sharding. For dense models, this includes all GEMM-side parameters assigned to the device group. For MoE models, it includes shared dense parameters and only the routed expert parameters used by the request, rather than all resident expert parameters. The factor of $2$ reflects one multiply-add per active parameter per request. This approximation captures the fact that GEMM-side arithmetic scales linearly with the number of concurrently processed requests.

\paragraph{GEMM-side memory traffic.}
For GEMM-dominated operators, we approximate the total memory traffic by the model-weight footprint:
\[
D^{\mathrm{gemm}}(b) \approx M_{\mathrm{weights}},
\]
where $M_{\mathrm{weights}}$ is the resident footprint of all model weights assigned to the modeled device group after sharding. In an MoE model, inactive experts do not contribute to $P_{\mathrm{act}}$ for an individual request, but their resident weights still contribute to $M_{\mathrm{weights}}$. The approximation $D^{\mathrm{gemm}}(b) \approx M_{\mathrm{weights}}$ corresponds to a sufficiently large batch whose routed tokens collectively access and amortize the resident expert-weight shards; the runtime planner uses measured operator costs rather than relying on this approximation.

\paragraph{Continuous form of the colocated batch-size limit.}
In the main text, the exact colocated feasible batch size is defined as
\[
b_{\max}(s)
=
\left\lfloor
\frac{M-M_{\mathrm{weights}}}{m_{\mathrm{kv}}(s)}
\right\rfloor.
\]
For the closed-form analysis, we use the continuous relaxation
\[
\bar b(s)
=
\frac{M-M_{\mathrm{weights}}}{m_{\mathrm{kv}}(s)}.
\]
Dropping the floor changes the feasible batch size by at most one request.

\paragraph{Derivation of the GEMM threshold.}
The threshold $b^*$ is defined as the batch size at which GEMM transitions from memory-bound to compute-bound execution. Under the approximations above, the GEMM memory-load time is
\[
\frac{D^{\mathrm{gemm}}(b)}{\beta}
\approx
\frac{M_{\mathrm{weights}}}{\beta},
\]
while the GEMM compute time is
\[
\frac{W^{\mathrm{gemm}}(b)}{F}
\approx
\frac{2bP_{\mathrm{act}}}{F}.
\]
Equating these two terms gives the transition point:
\[
\frac{M_{\mathrm{weights}}}{\beta}
=
\frac{2bP_{\mathrm{act}}}{F},
\]
and therefore
\[
b^* = \frac{M_{\mathrm{weights}}F}{2P_{\mathrm{act}}\beta}.
\]

\paragraph{Interpretation.}
When $b \ge b^*$, GEMM compute time exceeds weight-loading time, so GEMM operates in the compute-bound regime. When $b < b^*$, the opposite holds and GEMM is memory-bound. This threshold is central to the later gain analysis: homogeneous ODS begins to improve over colocation precisely when KV-cache pressure pushes the colocated feasible batch size below this compute-bound threshold.

\paragraph{Scope of the approximation.}
These approximations are introduced only to obtain interpretable closed-form characterizations of when ODS helps and what bounds its gain. They are not intended as a full performance model of the runtime system; the later planner and end-to-end evaluation use richer model- and hardware-specific information.

\subsection{Proofs for Section~\ref{sec:theory}}
\label{app:proofs}

This appendix provides complete proofs for the theorems stated in \Secref{sec:theory}.
We use the same notation and scaling approximations introduced in \Secref{sec:analytical_setup}: $D^{\mathrm{attn}}(b) = b \cdot m_{\mathrm{kv}}(s)$, $W^{\mathrm{gemm}}(b) = 2bP_{\mathrm{act}}$, $D^{\mathrm{gemm}}(b) = M_{\mathrm{weights}}$, and the continuous relaxation $\bar b(s) = (M - M_{\mathrm{weights}})/m_{\mathrm{kv}}(s)$.
For the homogeneous comparison, we restrict attention to sequence lengths with $b_{\max}(s) \ge 1$ and assume $m_{\mathrm{kv}}(s)$ is strictly increasing on this feasible domain, as stated in \Secref{sec:homo_vs_coloc}.

\subsubsection{Proof of Theorem~\ref{thm:homo_vs_coloc}: Homogeneous Disaggregation vs.\ Colocation}
\label{app:proof_homo}

\paragraph{Step 1: Simplify the disaggregated CPT}

Suppressing the fixed GPU-type index and expanding $\alpha=c/\beta$ and $\gamma=c/F$ in Eq.~\ref{eq:cpt_homo_disagg} gives:
\begin{equation}
    \label{eq:app_cpt_disagg}
    \mathrm{CPT}^{\mathrm{hom}} = \frac{c}{b}\left(\frac{b \cdot m_{\mathrm{kv}}(s)}{\beta} + \frac{2bP_{\mathrm{act}}}{F}\right) = c\left(\frac{m_{\mathrm{kv}}(s)}{\beta} + \frac{2P_{\mathrm{act}}}{F}\right).
\end{equation}
The batch size $b$ cancels because both the attention memory traffic and the GEMM compute scale linearly with $b$.
The disaggregated CPT depends on $s$ only through $m_{\mathrm{kv}}(s)$.

\paragraph{Step 2: Simplify the colocated CPT}

For an arbitrary local batch size $b$, the colocated cost derived in Appendix~\ref{app:cost-expression-details} becomes:
\begin{align}
    \mathrm{CPT}^{\mathrm{coloc}}(s,b)
    &= \frac{c}{b}\!\left(\frac{b \cdot m_{\mathrm{kv}}(s)}{\beta} + \max\!\left(\frac{M_{\mathrm{weights}}}{\beta},\; \frac{2bP_{\mathrm{act}}}{F}\right)\right) \nonumber \\
    &= c\!\left(\frac{m_{\mathrm{kv}}(s)}{\beta} + \frac{1}{b}\max\!\left(\frac{M_{\mathrm{weights}}}{\beta},\; \frac{2bP_{\mathrm{act}}}{F}\right)\right).
\end{align}
Here $b$ is a local batch-size variable; Section~\ref{sec:formulation} distinguishes the global, per-replica, and microbatch quantities used by the planner. The cost is non-increasing in $b$, so its minimum over $1 \le b \le \bar b(s)$ is attained at $b=\bar b(s)$, yielding $\mathrm{CPT}^{\mathrm{coloc}}(s)$ in Eq.~\ref{eq:cpt_coloc}.
The two GEMM regimes yield:

\emph{Compute-bound GEMM} ($\bar b(s) \geq b^*$):
\begin{equation}
    \label{eq:app_coloc_compute}
    \mathrm{CPT}^{\mathrm{coloc}} = c\left(\frac{m_{\mathrm{kv}}(s)}{\beta} + \frac{2P_{\mathrm{act}}}{F}\right).
\end{equation}

\emph{Memory-bound GEMM} ($\bar b(s) < b^*$):
\begin{equation}
    \label{eq:app_coloc_memory}
    \mathrm{CPT}^{\mathrm{coloc}} = c\left(\frac{m_{\mathrm{kv}}(s)}{\beta} + \frac{M_{\mathrm{weights}}}{\bar b(s) \cdot \beta}\right).
\end{equation}

\paragraph{Step 3: Case 1: Compute-bound GEMM ($\bar b(s) \geq b^*$)}

When the colocated batch size is large enough for GEMM to saturate compute, comparing Eq.~\ref{eq:app_cpt_disagg} and Eq.~\ref{eq:app_coloc_compute} gives
$$
\mathrm{CPT}^{\mathrm{coloc}} = c\left(\frac{m_{\mathrm{kv}}(s)}{\beta} + \frac{2P_{\mathrm{act}}}{F}\right) = \mathrm{CPT}^{\mathrm{hom}}.
$$
Therefore $G_{\mathrm{hom}}(s) = 1$.
This establishes claim~1: when the batch size is large enough for GEMM to fully utilize compute under colocation, there is no efficiency gap for disaggregation to exploit.

\paragraph{Step 4: Case 2: Memory-bound GEMM ($\bar b(s) < b^*$)}

When KV-cache pressure pushes $\bar b(s)$ below $b^*$, GEMM under colocation becomes memory-bound while disaggregation maintains compute-bound GEMM.

\subparagraph{Step 4a: Rewrite the colocated CPT.}
Substituting $\bar b(s) = (M - M_{\mathrm{weights}})/m_{\mathrm{kv}}(s)$ into Eq.~\ref{eq:app_coloc_memory}:
\begin{align}
    \mathrm{CPT}^{\mathrm{coloc}} &= c\left(\frac{m_{\mathrm{kv}}(s)}{\beta} + \frac{M_{\mathrm{weights}} \cdot m_{\mathrm{kv}}(s)}{(M - M_{\mathrm{weights}}) \cdot \beta}\right) \nonumber \\
    &= \frac{c \cdot m_{\mathrm{kv}}(s)}{\beta}\left(1 + \frac{M_{\mathrm{weights}}}{M - M_{\mathrm{weights}}}\right) \nonumber \\
    &= \frac{c \cdot m_{\mathrm{kv}}(s)}{\beta} \cdot \frac{M}{M - M_{\mathrm{weights}}}. \label{eq:app_coloc_simplified}
\end{align}

\subparagraph{Step 4b: Cost gain expression.}
Dividing Eq.~\ref{eq:app_coloc_simplified} by Eq.~\ref{eq:app_cpt_disagg}:
\begin{equation}
    \label{eq:app_gain_expr}
    G_{\mathrm{hom}}(s) = \frac{m_{\mathrm{kv}}(s)/\beta}{m_{\mathrm{kv}}(s)/\beta + 2P_{\mathrm{act}}/F} \cdot \frac{M}{M - M_{\mathrm{weights}}}.
\end{equation}

\subparagraph{Step 4c: Monotonicity.}
Let $x = m_{\mathrm{kv}}(s)/\beta$, which increases with $s$ since $m_{\mathrm{kv}}(s)$ is increasing in $s$.
Define $A = M/(M - M_{\mathrm{weights}}) > 1$ (since $M_{\mathrm{weights}} > 0$) and $\kappa = 2P_{\mathrm{act}}/F > 0$.
Then
\begin{equation}
    G_{\mathrm{hom}} = \frac{Ax}{x + \kappa}, \qquad \frac{dG_{\mathrm{hom}}}{dx} = \frac{A\kappa}{(x + \kappa)^2} > 0.
\end{equation}
Since $dG_{\mathrm{hom}}/dx > 0$ for all $x > 0$, the cost gain $G_{\mathrm{hom}}$ is strictly increasing in $x$ and therefore in $s$.
This establishes claim~2.

\subparagraph{Step 4d: Upper bound.}
For any feasible $s$ in the memory-bound regime, $G_{\mathrm{hom}}(s) < A$ because $Ax/(x + \kappa) < A$ for every finite $x > 0$.
Combining this fact with the compute-bound case, where $G_{\mathrm{hom}}(s) = 1 \leq A$, gives the following bound for every feasible $s$ with $b_{\max}(s) \ge 1$:
$$
1 \;\leq\; G_{\mathrm{hom}}(s) \;<\; \frac{M}{M - M_{\mathrm{weights}}}.
$$
This establishes claim~3 and completes the proof.
\hfill$\square$

\paragraph{Summary}
The cost gain is $G_{\mathrm{hom}}(s)=1$ when $\bar b(s) \geq b^*$.
When $\bar b(s) < b^*$, it is
\begin{equation}
    \label{eq:app_gain_summary}
    G_{\mathrm{hom}}(s)
    =
    \frac{m_{\mathrm{kv}}(s)/\beta}{m_{\mathrm{kv}}(s)/\beta + 2P_{\mathrm{act}}/F}
    \cdot
    \frac{M}{M - M_{\mathrm{weights}}}.
\end{equation}
Here, equality holds in the compute-bound case, while the displayed formula applies in the memory-bound case.
Over the feasible memory-bound domain, $G_{\mathrm{hom}}$ is strictly increasing in $s$ and bounded above by $M/(M - M_{\mathrm{weights}})$.

\subsubsection{Proof of Theorem~\ref{thm:hetero_vs_homo}: Heterogeneous vs.\ Homogeneous Disaggregation}
\label{app:proof_hetero}

Let $\mathcal{I}$ be the feasible interval from Theorem~\ref{thm:hetero_vs_homo}, and let $x=m_{\mathrm{kv}}(s)$. We first establish the shape of the analytically extended gain as a function of $x \ge 0$, then restrict it to $x \in m_{\mathrm{kv}}(\mathcal{I})$.

\paragraph{Step 1: $G_{\mathrm{het}}(s) \geq 1$ (Claims 1 and 2)}

For each GPU type $j$, the heterogeneous CPT satisfies
\begin{align}
    \mathrm{CPT}^{\mathrm{het}}_{\star}
    &= \min(\alpha_1, \alpha_2) \cdot m_{\mathrm{kv}}(s) + \min(\gamma_1, \gamma_2) \cdot 2P_{\mathrm{act}} \nonumber \\
    &\leq\; \alpha_j \cdot m_{\mathrm{kv}}(s) + \gamma_j \cdot 2P_{\mathrm{act}}
    \;=\; \mathrm{CPT}^{\mathrm{hom}}_j,
\end{align}
because $\min(\alpha_1, \alpha_2) \leq \alpha_j$ and $\min(\gamma_1, \gamma_2) \leq \gamma_j$ for all $j$.
Since this holds for every $j$, it holds for the minimum:
$$
\mathrm{CPT}^{\mathrm{het}}_{\star} \leq \min_j\, \mathrm{CPT}^{\mathrm{hom}}_j = \mathrm{CPT}^{\mathrm{hom}}_{\star}.
$$
Therefore $G_{\mathrm{het}}(s) \geq 1$ for every $s \in \mathcal{I}$, establishing claim~1.

For claim~2: if GPU type $j$ dominates type $k$ on both metrics ($\alpha_j \leq \alpha_k$ and $\gamma_j \leq \gamma_k$), then $\min(\alpha_1, \alpha_2) = \alpha_j$ and $\min(\gamma_1, \gamma_2) = \gamma_j$.
Hence $\mathrm{CPT}^{\mathrm{het}}_{\star} = \mathrm{CPT}^{\mathrm{hom}}_j = \mathrm{CPT}^{\mathrm{hom}}_{\star}$ and $G_{\mathrm{het}}(s) = 1$ for every $s \in \mathcal{I}$.

\paragraph{Step 2: Unimodal structure (Claim 3)}

When neither type dominates, the metric orderings must be crossed. Under the theorem's labeling $\eta \ge 1$, the only possible crossed ordering is $\alpha_1 < \alpha_2$ and $\gamma_2 < \gamma_1$; the reverse ordering would imply $\eta < 1$.
Define $r = \alpha_2/\alpha_1 > 1$ and $q = \gamma_1/\gamma_2 > 1$.
The cost expressions become:
\begin{align}
    \mathrm{CPT}^{\mathrm{hom}}_1 &= \alpha_1 \cdot m_{\mathrm{kv}}(s) + q\gamma_2 \cdot 2P_{\mathrm{act}}, \label{eq:app_homo1} \\
    \mathrm{CPT}^{\mathrm{hom}}_2 &= r\alpha_1 \cdot m_{\mathrm{kv}}(s) + \gamma_2 \cdot 2P_{\mathrm{act}}, \label{eq:app_homo2} \\
    \mathrm{CPT}^{\mathrm{het}}_{\star} &= \alpha_1 \cdot m_{\mathrm{kv}}(s) + \gamma_2 \cdot 2P_{\mathrm{act}}. \label{eq:app_hetero}
\end{align}

\subparagraph{Step 2a: Crossing point.}
The two homogeneous CPTs are equal when
\begin{align*}
\alpha_1 \cdot m_{\mathrm{kv}}(s) + q\gamma_2 \cdot 2P_{\mathrm{act}}
&= r\alpha_1 \cdot m_{\mathrm{kv}}(s) + \gamma_2 \cdot 2P_{\mathrm{act}},
\end{align*}
which gives the crossing traffic
\begin{equation}
    \label{eq:app_crossing}
    x^* = \frac{(q - 1)\,\gamma_2 \cdot 2P_{\mathrm{act}}}{(r - 1)\,\alpha_1}
    = \frac{2P_{\mathrm{act}}(\gamma_1-\gamma_2)}{\alpha_2-\alpha_1}.
\end{equation}
For $x < x^*$, GPU type~2 is the better homogeneous choice ($\mathrm{CPT}^{\mathrm{hom}}_2 < \mathrm{CPT}^{\mathrm{hom}}_1$).
For $x > x^*$, GPU type~1 is the better homogeneous choice.

\subparagraph{Step 2b: Monotonicity in each sub-region.}
Let $x = m_{\mathrm{kv}}(s)$ for brevity.

\emph{Sub-region 1} ($x \leq x^*$, short sequences, homogeneous uses GPU type~2):
\begin{equation}
    G_{\mathrm{het}}(x) = \frac{r\alpha_1 x + \gamma_2 \cdot 2P_{\mathrm{act}}}{\alpha_1 x + \gamma_2 \cdot 2P_{\mathrm{act}}}\,.
\end{equation}
Differentiating,
\begin{equation}
    \frac{dG_{\mathrm{het}}}{dx} = \frac{(r - 1)\,\alpha_1 \cdot \gamma_2 \cdot 2P_{\mathrm{act}}}{(\alpha_1 x + \gamma_2 \cdot 2P_{\mathrm{act}})^2} > 0.
\end{equation}
So $G_{\mathrm{het}}$ is strictly increasing in this sub-region.
At $x = 0$: $G_{\mathrm{het}}(0) = (\gamma_2 \cdot 2P_{\mathrm{act}})/(\gamma_2 \cdot 2P_{\mathrm{act}}) = 1$.

\emph{Sub-region 2} ($x \geq x^*$, long sequences, homogeneous uses GPU type~1):
\begin{equation}
    G_{\mathrm{het}}(x) = \frac{\alpha_1 x + q\gamma_2 \cdot 2P_{\mathrm{act}}}{\alpha_1 x + \gamma_2 \cdot 2P_{\mathrm{act}}}\,.
\end{equation}
Differentiating,
\begin{equation}
    \frac{dG_{\mathrm{het}}}{dx} = \frac{-(q - 1)\,\alpha_1 \cdot \gamma_2 \cdot 2P_{\mathrm{act}}}{(\alpha_1 x + \gamma_2 \cdot 2P_{\mathrm{act}})^2} < 0.
\end{equation}
So $G_{\mathrm{het}}$ is strictly decreasing in this sub-region.
As $x \to \infty$: $G_{\mathrm{het}} \to \alpha_1/\alpha_1 = 1$.

Combining both sub-regions, the analytic gain $G_{\mathrm{het}}(x)$ increases from $1$ at $x=0$, peaks at $x=x^*$, and decreases toward $1$ as $x \to \infty$. Because $m_{\mathrm{kv}}(s)$ is continuous and strictly increasing on $\mathcal{I}$, if $x^* \in m_{\mathrm{kv}}(\mathcal{I})$, there is a unique $s^* \in \mathcal{I}$ satisfying $m_{\mathrm{kv}}(s^*)=x^*$ and the restricted gain peaks there. If $x^*$ lies outside $m_{\mathrm{kv}}(\mathcal{I})$, the feasible interval contains only one side of the analytic curve, so the restricted gain is monotone. This establishes claim~3.

\paragraph{Step 3: Upper-bound part of Claim 3}

By the unimodality established in Step~2, the supremum of the analytic gain over $x \ge 0$ is attained at the crossing point $x=x^*$. The same upper bound therefore applies to any restricted feasible interval.

\subparagraph{Step 3a: Peak gain at $x^*$.}
Using the sub-region~1 expression at $x=x^*$ and noting from Eq.~\ref{eq:app_crossing} that $\alpha_1 x^* = (q-1)\gamma_2 \cdot 2P_{\mathrm{act}}/(r-1)$:
\begin{align}
    G_{\mathrm{het}}(x^*) &= \frac{r\alpha_1 x^* + \gamma_2 \cdot 2P_{\mathrm{act}}}{\alpha_1 x^* + \gamma_2 \cdot 2P_{\mathrm{act}}}
    = \frac{r \cdot \tfrac{(q-1)}{(r-1)} + 1}{\tfrac{(q-1)}{(r-1)} + 1} \nonumber \\
    &= \frac{r(q-1) + (r-1)}{(q-1) + (r-1)}
    = \frac{rq - 1}{r + q - 2}. \label{eq:app_peak}
\end{align}

\subparagraph{Step 3b: Factor out cost ratio.}
Let $w = c_1/c_2$.
Then $r = \alpha_2/\alpha_1 = \beta_1/(w\,\beta_2)$ and $q = \gamma_1/\gamma_2 = w\,F_2/F_1$.
The product is cost-independent:
\begin{equation}
    rq = \frac{\beta_1\,F_2}{\beta_2\,F_1} = \eta.
\end{equation}
The peak gain $G_{\mathrm{het}}(x^*) = (\eta - 1)/(r + q - 2)$ is decreasing in $r + q$ for fixed $rq = \eta > 1$.
To obtain the tightest bound, we minimize $r + q$ subject to $rq = \eta$:
$$
r + q = \frac{\beta_1}{w\,\beta_2} + \frac{w\,F_2}{F_1}.
$$
By the AM-GM inequality, $r + q \geq 2\sqrt{rq} = 2\sqrt{\eta}$, with equality when $r = q = \sqrt{\eta}$, which occurs at the cost ratio
$$
w^* = \sqrt{\frac{\beta_1 F_1}{\beta_2 F_2}}.
$$

\subparagraph{Step 3c: Evaluate the bound.}
Substituting $r + q = 2\sqrt{\eta}$ and $rq = \eta$:
\begin{align}
    \sup_{x \ge 0,\,w}\; G_{\mathrm{het}}(x)
    &= \frac{\eta - 1}{2\sqrt{\eta} - 2}
    = \frac{(\sqrt{\eta} - 1)(\sqrt{\eta} + 1)}{2(\sqrt{\eta} - 1)} \nonumber \\
    &= \frac{1 + \sqrt{\eta}}{2}.
\end{align}
This establishes the upper-bound part of claim~3 and completes the proof.
\hfill$\square$

\paragraph{Summary}
The cost gain in the non-trivial case ($\alpha_1 < \alpha_2$, $\gamma_2 < \gamma_1$) can be expressed as
\begin{equation}
    \label{eq:app_hetero_summary}
    G_{\mathrm{het}}(s) = \begin{cases}
    \dfrac{r\alpha_1 m_{\mathrm{kv}}(s) + \gamma_2 \cdot 2P_{\mathrm{act}}}{\alpha_1 m_{\mathrm{kv}}(s) + \gamma_2 \cdot 2P_{\mathrm{act}}} & \mathrm{if}\; m_{\mathrm{kv}}(s) \leq x^*, \\[10pt]
    \dfrac{\alpha_1 m_{\mathrm{kv}}(s) + q\gamma_2 \cdot 2P_{\mathrm{act}}}{\alpha_1 m_{\mathrm{kv}}(s) + \gamma_2 \cdot 2P_{\mathrm{act}}} & \mathrm{if}\; m_{\mathrm{kv}}(s) \geq x^*.
    \end{cases}
\end{equation}
The analytic gain is unimodal in $x=m_{\mathrm{kv}}(s)$, with $G_{\mathrm{het}}(0) = 1$, a peak of $(rq - 1)/(r + q - 2)$ at $x=x^*$, and $G_{\mathrm{het}}(x) \to 1$ as $x \to \infty$. On the feasible interval $\mathcal{I}$, this peak is attained only when $x^* \in m_{\mathrm{kv}}(\mathcal{I})$; otherwise, the restricted gain is monotone.
The tightest upper bound over all non-negative traffic values and all cost ratios is
$$
\sup_{x \ge 0,\,w}\; G_{\mathrm{het}}(x) = \frac{1 + \sqrt{\beta_1 F_2 / (\beta_2 F_1)}}{2}.
$$

\subsection{CAD Feasibility Condition for Uniform-Layer Models}
\label{app:cad_feasibility_condition}

This section gives a simple condition for the feasibility mismatch discussed in \Secref{sec:cad_limits}.
The condition applies to the uniform-layer CAD setting, where every layer has the same attention pattern and therefore the attention side can be described by a single feasible latency range.
Hybrid-attention architectures, where full-attention, sliding-window-attention, or linear-attention layers may have different latency and memory behavior, require the separate discussion in \Secref{sec:cad_limits}.

Following the heterogeneous notation in \Secref{sec:analytical_setup}, let GPU type~1 serve the attention side and GPU type~2 serve the rest-of-model side.
Let $M_1$ be the memory capacity of one GPU of type~1, $\beta_1$ be its memory bandwidth, and $\mu$ be the number of micro-batches sharing the attention-side memory.
Then a simple upper bound on the feasible attention-stage latency is
\begin{equation}
    \label{eq:cad_attn_latency_max}
    T_{\mathrm{attn}}^{\max}
    =
    \frac{M_1}{\mu\beta_1}.
\end{equation}
This expression reflects that each micro-batch can use at most $M_1 / \mu$ bytes of attention-side memory, which can be streamed at bandwidth $\beta_1$.

Let $\pi_2$ denote the parallelism strategy on the rest-of-model side.
Let $M_{\mathrm{weights},2}(\pi_2)$ be the per-GPU model-weight footprint on GPU type~2 under this strategy, and let $\beta_2$ be the memory bandwidth of GPU type~2.
The rest-of-model stage must at least load its shard of the model weights, so its latency is lower-bounded by
\begin{equation}
    \label{eq:cad_rest_latency_min}
    T_{\mathrm{rest}}^{\min}(\pi_2)
    =
    \frac{M_{\mathrm{weights},2}(\pi_2)}{\beta_2}.
\end{equation}
For example, under an even sharding strategy over $g_2(\pi_2)$ GPUs, one can approximate
\begin{equation}
    \label{eq:cad_weight_shard}
    M_{\mathrm{weights},2}(\pi_2)
    \approx
    \frac{M_{\mathrm{weights}}}{g_2(\pi_2)}.
\end{equation}

For CAD to be feasible under this simplified uniform-layer model, the feasible latency ranges of the two stages must overlap.
A necessary condition is therefore
\begin{equation}
    \label{eq:cad_feasibility_condition}
    T_{\mathrm{attn}}^{\max}
    \geq
    T_{\mathrm{rest}}^{\min}(\pi_2),
    \qquad\mathrm{i.e.,}\qquad
    \frac{M_1}{\mu\beta_1}
    \geq
    \frac{M_{\mathrm{weights},2}(\pi_2)}{\beta_2}.
\end{equation}
When this inequality fails, even the largest feasible attention-stage latency is smaller than the unavoidable weight-loading latency on the rest-of-model side.
In that case, the two CAD stages cannot be balanced, and pipeline bubbles are unavoidable.

\section{Detailed Serving-Plan Formulation}
\label{app:detailed-formulation}

This appendix section expands the compact formulation in \Secref{sec:formulation}.
The main text keeps only the abstractions needed to motivate the planner; here
we spell out the structural definitions, batching notation, execution and
communication latency, and memory constraint.

\subsection{Model Representation and Partition}
\label{sec:model-partition}

We model one decode step as a computation DAG
$\mathcal{G}=(\mathcal{V},\mathcal{E})$, where each node
$v\in\mathcal{V}$ is an operator and each edge $(u,v)\in\mathcal{E}$ is a data
dependency.  During serving, operators are executed according to a fixed
topological order.  For transformer inference this order is effectively a
linear operator sequence
\[
  \mathbf{o}=(o_1,o_2,\ldots,o_{N_{\mathrm{model}}}),
\]
and all partitioning decisions below refer to positions in this sequence.

Let $L_{\mathrm{model}}$ be the number of transformer layers.  A layer
boundary is the first operator of a layer; we denote these positions by
$\ell_1<\ell_2<\cdots<\ell_{L_{\mathrm{model}}}$, with $\ell_1=1$.  A
\textbf{partition block} is the minimal layer-aligned repeating unit of the
operator sequence.  Formally, it spans $L_{\mathrm{block}}$ consecutive layers,
or equivalently $N_{\mathrm{block}}$ consecutive operators, where
$L_{\mathrm{block}}$ is the smallest positive period such that the operator
pattern of layers
$\ell_k,\ldots,\ell_{k+L_{\mathrm{block}}-1}$ repeats for every block-aligned
starting layer~$k$.  We take the first partition block to start at
$\ell_1$ and assume $L_{\mathrm{block}}$ divides $L_{\mathrm{model}}$, giving
$B_{\mathrm{total}}=L_{\mathrm{model}}/L_{\mathrm{block}}$ partition blocks.

Within one partition block, the optimizer chooses a stage template
\[
  \mathbf{q}=(q_0,q_1,\ldots,q_S),
  \qquad
  0=q_0<q_1<\cdots<q_S=N_{\mathrm{block}} .
\]
Stage~$j$ contains operators at relative positions
$q_{j-1}+1,\ldots,q_j$ inside the block.  The same template is applied to
every partition block in the model, so stage~$j$ denotes the same relative
operator slice in each repeated block.

\subsection{Device-Group Types, Replicas, and Placement}
\label{sec:placement}

A device-group type~$m$ is a logical configuration $(h_m,\pi_m)$, where
$h_m$ is the GPU type and $\pi_m$ is the intra-group parallelism strategy,
such as tensor parallelism, expert parallelism, or their composition.  A
device-group replica is one physical instance of that type.  Replicas of the
same type have identical configuration and execute the same stage set, but
serve disjoint subsets of the global resident batch.

We restrict the number of stages to $S=kK$ for an integer $k\ge 1$.  Stages are
assigned to the ordered device-group types by the round-robin map
\begin{equation}
\label{eq:placement}
  \sigma(j) = ((j-1)\bmod K)+1,
  \qquad j=1,\ldots,S .
\end{equation}
This map is applied identically within every partition block.  Consequently,
each type owns exactly $k$ stage positions per block, and the stage-to-type
assignment repeats with the same partition template.
The stage set assigned to type~$m$ is therefore
\[
  \mathcal{S}_m=\{j:1\le j\le S,\ \sigma(j)=m\}.
\]
Every replica of type~$m$ executes all stages in $\mathcal{S}_m$.

Let $B$ be the global resident batch size across the full deployment.  If a
replica of type~$m$ has resident batch size $b_m$, then the number of replicas
of that type is
\begin{equation}
\label{eq:replica-count}
  n_m = \frac{B}{b_m},
  \qquad m=1,\ldots,K ,
\end{equation}
which requires $b_m$ to divide $B$.  This coupling ensures that all
device-group types collectively process the same global batch.

\subsection{Scheduling Parameters}
\label{sec:scheduling}

The resident batch on each replica is split into $\mu$ microbatches.  We write
\[
  \widehat{B}=\frac{B}{\mu},
  \qquad
  \widehat{b}_m=\frac{b_m}{\mu},
\]
for the global microbatch size and per-replica microbatch size, respectively.
The divisibility requirements are $\mu\mid B$ and $\mu\mid b_m$ for all
device-group types.  The resident batch~$b_m$ determines the number of tokens
completed by a replica per round trip, while $\widehat{b}_m$ determines the
work executed in one pipeline scheduling step.

The execution latency of stage~$j$ is
\begin{equation}
\label{eq:stage-lat}
  t_j =
  f_{\mathrm{lat}}\!\left(
    \mathbf{o}_{q_{j-1}+1:q_j},
    (h_{\sigma(j)},\pi_{\sigma(j)}),
    \widehat{b}_{\sigma(j)}
  \right),
  \qquad j=1,\ldots,S .
\end{equation}
Let $A_j$ denote the activation interface at the boundary after stage~$j$,
including its tensor shape, byte size per request, and parallel layout.  Define
$j^+=j+1$ for $j<S$ and $j^+=1$ for $j=S$; the latter is the boundary between
consecutive partition blocks.  The corresponding transfer latency is
\begin{equation}
\label{eq:comm-lat}
\begin{split}
  u_j = f_{\mathrm{comm}}\!\big(&A_j,
    h_{\sigma(j)},\pi_{\sigma(j)},\widehat{b}_{\sigma(j)},\\
    &h_{\sigma(j^+)},\pi_{\sigma(j^+)},
    \widehat{b}_{\sigma(j^+)}\big),
\end{split}
\end{equation}
where $f_{\mathrm{comm}}$ is obtained from communication profiles and accounts
for any resharding or redistribution between the source and destination
replicas.  It returns the local handoff cost when the adjacent stages execute
on the same replica.  The $j=S$ transfer occurs between consecutive blocks and
is omitted after the final block.

The steady-state round-trip time is then
\begin{equation}
\label{eq:rtt}
  T_{\mathrm{rtt}}
  =
  f_{\mathrm{rtt}}(\mu,B_{\mathrm{total}},
  \mathbf{t},\mathbf{u}),
\end{equation}
where $\mathbf{t}=(t_1,\ldots,t_S)$ and
$\mathbf{u}=(u_1,\ldots,u_S)$.  The scheduling model enforces execution and
transfer dependencies, models contention for compute and communication
resources, and permits overlap only when the selected schedule and resources
allow it.  The planner evaluates this function by lightweight simulation.

\subsection{Memory Constraint}
\label{sec:constraints}

For each device-group type~$m$, $M_{\mathrm{weights},m}$ denotes the weight
memory for the stages assigned to that type, including all repeated partition
blocks hosted by a replica.  Let
$\mathrm{KV}_m(\widehat{b}_m,s)$ denote the KV-cache memory required by one
per-replica microbatch at sequence length~$s$; this term is zero for stages
that do not own KV-resident attention state.  Since all $\mu$ microbatches are
resident in the pipeline, a replica of type~$m$ must satisfy
\[
  M_{\mathrm{weights},m}
  + \mu\,\mathrm{KV}_m(\widehat{b}_m,s)
  \le M_m ,
\]
where $M_m$ is the effective memory capacity of the device group after
accounting for the parallelism strategy~$\pi_m$.  Together with the SLO
constraint and the objective in \eqref{eq:objective}, this yields the compact
optimization problem in \eqref{eq:full-opt}.  The cost-per-token objective is
derived from global batch size and replica counts in
Appendix~\ref{app:cost-derivation}.

\subsection{Derivation of the Cost-per-Token Objective}
\label{app:cost-derivation}

We derive the general serving-plan objective~$\mathrm{CPT}_{\mathrm{plan}}$
in~\eqref{eq:objective} from the
perspective of the global batch size and device-group replicas introduced in
Section~\ref{sec:placement}.

\paragraph{Setup.}
Consider a global resident batch of $B$ requests processed in parallel across
all replicas.  Each replica of device-group type~$m$ holds a per-replica
resident batch of $b_m$ requests.  When the schedule uses $\mu$ micro-batches,
the global micro-batch size is $\widehat{B}=B/\mu$ and the per-replica
micro-batch size is $\widehat{b}_m=b_m/\mu$.  The system therefore requires
\begin{equation}
\label{eq:replica-derive}
  n_m \;=\; \frac{B}{b_m}
  \;=\; \frac{\widehat{B}}{\widehat{b}_m}
\end{equation}
replicas of type~$m$.

\paragraph{Throughput.}
In steady-state pipeline execution, each replica completes one decoding step
(i.e., generates one output token for each of its $b_m=\mu\widehat{b}_m$
resident requests) every $T_{\mathrm{rtt}}$ seconds.  Since there are
$n_m = B/b_m$ replicas of type~$m$, and every request passes through every
device-group type, the system-wide throughput is determined by the global
resident batch:
\begin{equation}
\label{eq:throughput}
  \mathrm{Throughput}
  \;=\; \frac{B}{T_{\mathrm{rtt}}}
  \qquad \mathrm{tokens/s}.
\end{equation}

\paragraph{Total cost rate.}
The instantaneous monetary cost rate of operating all replicas is
\begin{equation}
\label{eq:total-cost-rate}
  \mathrm{CostRate}_{\mathrm{total}}
  \;=\; \sum_{m=1}^{K} n_m \cdot c_m
  \;=\; \sum_{m=1}^{K} \frac{B}{b_m}\,c_m
  \;=\; B\,\sum_{m=1}^{K} \frac{c_m}{b_m}.
\end{equation}

\paragraph{Cost per token.}
Dividing the total cost rate by the throughput yields
\begin{equation}
\label{eq:cost-per-token-derive}
\begin{aligned}
  \mathrm{CPT}_{\mathrm{plan}}
  &\;=\; \frac{\mathrm{CostRate}_{\mathrm{total}}}{\mathrm{Throughput}}
   \;=\; \frac{B\,\displaystyle\sum_{m=1}^{K} \frac{c_m}{b_m}}
              {\dfrac{B}{T_{\mathrm{rtt}}}} \\
  &\;=\; T_{\mathrm{rtt}} \;\cdot\; \sum_{m=1}^{K} \frac{c_m}{b_m},
\end{aligned}
\end{equation}
which recovers~\eqref{eq:objective}.  Note that the global batch size~$B$
cancels, confirming that the cost-per-token objective depends only on the
device-group configurations and per-replica resident batch sizes, equivalently
on $\mu\widehat{b}_m$, not on the absolute scale of the deployment.  This
cancellation assumes scale-invariant replication: adding replicas repeats the
same per-replica execution and transfer pattern.

\section{Planner Details for Section~\ref{sec:algorithms}}
\label{app:planner-details}

This appendix details the latency and cost lower bounds used by the
branch-and-bound search in \Secref{sec:frontier-branch-bound}.  These bounds
allow the planner to prune partial assignments before invoking the schedule
simulator.

\subsection{Branch-and-Bound Lower Bounds}
\label{app:bnb-bounds}

After selecting a prefix
$P_i=(p_1,\ldots,p_i)$ of owner points, the planner computes an optimistic
completion using the best remaining coordinate from each unassigned frontier.

For a point~$p$ of owner~$m$, the mandatory serialized compute work is
$C_m(p)=\mu\sum_{j\in\mathcal{S}_m}t_j(p)$.
The compute lower bound for a prefix is therefore
\begin{equation}
\label{eq:compute-lower-bound}
  L_{\mathrm{comp}}(P_i)
  =\max\left\{
      \max_{m\le i} C_m(p_m),
      \max_{m>i}\min_{p\in\mathcal{F}_m} C_m(p)
    \right\}.
\end{equation}
The planner similarly lower-bounds communication by accumulating mandatory
serialized transfer work at each network endpoint.  A boundary whose endpoint
has not yet been assigned is minimized over that owner's feasible replica
counts.  Boundaries are minimized independently, which can only make the
estimate optimistic.  Denoting this endpoint bound by
$L_{\mathrm{net}}(P_i)$ gives the valid latency lower bound
\begin{equation}
\label{eq:tpot-lower-bound}
  L_T(P_i)=\max\bigl\{L_{\mathrm{comp}}(P_i),
                         L_{\mathrm{net}}(P_i)\bigr\}
  \le T_{\mathrm{rtt}}.
\end{equation}
Suffix minima also lower-bound the normalized cost of any completion; let
$R_{\mathrm{LB}}(P_i)$ denote this bound.  The best possible objective below
the prefix satisfies
$\mathrm{CPT}_{\mathrm{LB}}(P_i)=L_T(P_i)\,R_{\mathrm{LB}}(P_i)
\le \mathrm{CPT}_{\mathrm{plan}}$.

\onecolumn
\section{Notation}
\label{app:notation}

\begin{table}[!ht]
\centering
\papertablestyle
\renewcommand{\arraystretch}{0.94}
\begin{tabular}{@{}l|p{0.8\linewidth}@{}}
\toprule
\multicolumn{1}{@{}c|}{\textbf{Symbol}} & \multicolumn{1}{c@{}}{\textbf{Description}} \\
\midrule
\multicolumn{2}{@{}l}{\textit{Model and workload}} \\[2pt]
$\mathbf{o}$                & Operator sequence $(o_1,\ldots,o_{N_{\mathrm{model}}})$ of one decode step \\
$N_{\mathrm{model}}$            & Total number of operators \\
$L_{\mathrm{model}}$            & Total number of transformer layers \\
$P_{\mathrm{act}}$          & GEMM-side parameters activated per request after model-parallel sharding (routed experts only for MoE) \\
$M_{\mathrm{weights}}$          & Resident model-weight footprint after sharding, including inactive experts \\
$D^{\mathrm{attn}}(b),\,D^{\mathrm{gemm}}(b)$ & Aggregated attention and GEMM memory traffic at batch size $b$ \\
$W^{\mathrm{gemm}}(b)$      & Aggregated GEMM FLOPs at batch size $b$ \\
$s$                         & Average sequence length (workload parameter) \\
$m_{\mathrm{kv}}(s)$            & Per-request KV-cache memory at sequence length $s$ \\
$b^*$                       & GEMM compute-bound threshold batch size \\
$b_{\max}(s)$               & Max feasible batch size under memory at sequence length $s$ \\
$\bar b(s)$                 & Continuous relaxation of $b_{\max}(s)$ used in closed-form analysis \\
\midrule
\multicolumn{2}{@{}l}{\textit{GPU hardware (per GPU type $j$)}} \\[2pt]
$c_j$                       & Unit-time monetary cost of GPU type $j$ (\$/s); shorthand $c$ in the single-type case, $c_1,c_2$ for two types \\
$\beta_j$                   & Memory bandwidth (bytes/s); shorthand $\beta$, or $\beta_1,\beta_2$ \\
$F_j$                       & Peak compute throughput (FLOPS); shorthand $F$, or $F_1,F_2$ \\
$M_j$                       & Memory capacity (bytes); shorthand $M$, or $M_1,M_2$ \\
$\alpha_j,\,\gamma_j$       & Monetary cost per byte of memory traffic and per FLOP, respectively, for GPU type $j$ \\
$\mathrm{CPT}$              & Cost per token; $\mathrm{CPT}_{\mathrm{plan}}$ is the serving-plan objective; superscripts $\mathrm{coloc}$/$\mathrm{hom}$/$\mathrm{het}$ denote serving modes \\
\midrule
\multicolumn{2}{@{}l}{\textit{Theoretical analysis (\Secref{sec:theory})}} \\[2pt]
$G_{\mathrm{hom}}(s)$       & Cost gain of homogeneous ODS over colocated serving \\
$G_{\mathrm{het}}(s)$       & Cost gain of heterogeneous over homogeneous ODS \\
$\eta$                      & Hardware ratio $(\beta_1 F_2)/(\beta_2 F_1)$ of two GPU types \\
\midrule
\multicolumn{2}{@{}l}{\textit{Model structure (partition block; \Secref{sec:formulation})}} \\[2pt]
$L_{\mathrm{block}}$            & Number of transformer layers per partition block \\
$N_{\mathrm{block}}$            & Number of operators per partition block \\
$B_{\mathrm{total}}$            & Number of partition blocks ($=L_{\mathrm{model}}/L_{\mathrm{block}}$) \\
\midrule
\multicolumn{2}{@{}l}{\textit{Decision variables (\Secref{sec:formulation})}} \\[2pt]
$S$                         & Number of stages per partition block \\
$\mathbf{q}$                & Stage-boundary offsets $(q_0,\ldots,q_S)$ within a block \\
$K$                         & Number of device-group types \\
$(h_m,\pi_m)$               & Device-group configuration (GPU type, parallelism) of type $m$ \\
$b_m$                       & Per-replica resident batch size of device-group type $m$ \\
$\widehat{b}_m$             & Per-replica micro-batch size of device-group type $m$ ($=b_m/\mu$) \\
$\mu$                       & Number of micro-batches \\
\midrule
\multicolumn{2}{@{}l}{\textit{Derived / scheduling quantities (\Secref{sec:formulation})}} \\[2pt]
$\mathcal{S}_m$             & Stages assigned to device-group type $m$ (placement group) \\
$\sigma(j)$                 & Placement map from stage $j$ to its device-group type \\
$B$                         & Global resident batch size across all replicas \\
$n_m$                       & Replicas of device-group type $m$ ($=B/b_m$) \\
$c_m$                       & Unit-time monetary cost of one replica of device-group type $m$ \\
$M_m$                       & Effective memory capacity of a replica of type $m$ under $\pi_m$ \\
$M_{\mathrm{weights},m}$      & Total weight memory of stages assigned to type $m$ \\
$\mathrm{KV}_m(\widehat{b}_m,s)$ & KV-cache memory for one per-replica micro-batch on a replica of type $m$ \\
$t_j$                       & Execution latency of stage $j$ \\
$u_j$                       & Activation-transfer latency at the boundary after stage $j$ \\
$T_{\mathrm{rtt}}$              & Pipeline round-trip time (per-token latency in steady state) \\
$T_{\mathrm{SLO}}$              & TPOT service-level objective \\
\midrule
\multicolumn{2}{@{}l}{\textit{Planner (\Secref{sec:algorithms})}} \\[2pt]
$L_{\mathrm{sub}}$, $n$     & Sub-block length in layers and repetitions per block ($n=L_{\mathrm{block}}/L_{\mathrm{sub}}$) \\
$\mathbf{q}_{\mathrm{sub}}$ & Sub-block stage template, tiled over the $n$ repetitions \\
$\mathbf{T}_m(p)$           & Stage-latency vector of owner $m$'s point $p$ over $\mathcal{S}_m$ \\
$r_m(p),\,\rho_m(p)$        & Normalized cost and GPU count of point $p$: $c_m(p)/b_m(p)$, $g_m(p)/b_m(p)$ \\
$\mathcal{F}_m$             & Network-safe Pareto frontier of owner $m$ \\
\bottomrule
\end{tabular}
\caption{Unified notation used throughout the paper.}
\label{tab:notation}
\end{table}

\end{document}